\documentclass[aps,twocolumn,showpacs,amsmath,amssymb,pra,superscriptaddress,floatfix,longbibliography]{revtex4-1}
\usepackage{braket}
\usepackage{amsmath}
\usepackage{amssymb}
\usepackage{mathtools}
\usepackage{graphicx}
\usepackage{dcolumn}
\usepackage{bm}
\usepackage{multirow}
\usepackage{leftidx}
\usepackage{color}
\usepackage{float}
\usepackage{xcolor}
\usepackage{soul}

\usepackage{amsfonts}
\usepackage{eufrak}
\usepackage[german,english]{babel}

\begin{document}
\title{Intensity effects of light coupling to one- or two-atom
arrays of infinite extent}
\author{F.~Robicheaux}
\email{robichf@purdue.edu}
\affiliation{Department of Physics and Astronomy, Purdue University, West Lafayette,
Indiana 47907, USA}
\affiliation{Purdue Quantum Science and Engineering Institute, Purdue
University, West Lafayette, Indiana 47907, USA}
\author{Deepak A.~Suresh}
\affiliation{Department of Physics and Astronomy, Purdue University, West Lafayette,
Indiana 47907, USA}

\date{\today}

\begin{abstract}
We theoretically and computationally investigate the behavior of infinite atom
arrays when illuminated by nearly resonant light. 
We use higher order mean field equations to
investigate the coherent reflection and transmission and incoherent
scattering of photons from a single array and from a pair of
arrays as a function of
detuning for different values of the Rabi frequency. 
For the single array case, we show how increasing the light intensity changes
the probabilities for these different processes. For example, the incoherent
scattering probability initially increases with light intensity before 
decreasing at higher values.
For a pair of parallel arrays at near resonant separation, the effects from
increasing light intensity can become apparent with incredibly low intensity
light.
In addition, we derive the higher order mean field equations for these
infinite arrays giving a representation that can be evaluated with a finite
number of equations.

\end{abstract}

\maketitle

\section{Introduction}

Interesting many body effects occur when light interacts with many
atoms with separations less than the light's wavelength. In this
case, a photon coherently interacts with many atoms leading
to collective effects that can be difficult to anticipate from
single atom-photon interactions. There are
several interesting scenarios when the atoms are in a regular array
\cite{CYL2004,MSS2014,JR12012,BGA2016,SR22016,FJR2016,SWL2017,GGV2018,CDG2018,
AMA2017,AKC2019,HDC2019,NLO2019,QR12019,ZM12019,MFO2020,JR12019,
WBR2020,WR12020,BLG2020,GGV2019,
BR12020,BPP2020,RWR2020,CES2020,BR12021,DC12021,BR12022,PZP2023,
PWS2021,FR12021,RBY2022,MAG2022,SMA2022,ROY2023,ROYb2023}
because the dephasing that occurs due to the dipole-dipole interaction
is reduced and photon interference from different emitters can
lead to qualitatively new phenomena. One
fascinating scenario involves the manipulation of light using
atom arrays\cite{JR12012,BGA2016,SR22016,FJR2016,SWL2017,GGV2018,
JR12019,WBR2020,WR12020,GGV2019,RWR2020,BLG2020,
BR12020,BPP2020,CES2020,BR12021,DC12021,BR12022,PZP2023}
by modifying transmission, reflection, scattering, and diffraction.

Most previous studies have investigated atomic arrays with much
less than 1 excitation on average; the exceptions we have noted are in
Refs.~\cite{QR12019,ZM12019,MFO2020,WBR2020,WR12020,BLG2020,DC12021,PZP2023,FR12021,RBY2022,MAG2022,SMA2022,ROY2023,ROYb2023}. This limit holds when very weak light interacts with atoms
initially in the ground state. We denote this
limit of very weak light as the weak field approximation (WFA). In this
limit, the atoms can be treated as harmonic oscillators instead of
2-level systems which means most results are indistinguishable from
the interaction of light with classical oscillators. Although classical
electromagnetism explains most of these results, several groups have
proposed or measured quite interesting phenomena.
At the other limit
of computational complexity are calculations that utilize the full
density matrix as in
Refs.~\cite{QR12019,ZM12019,MFO2020,WBR2020,WR12020,PZP2023,FR12021,MAG2022,SMA2022}. For many systems, a mean-field approximation would lead to
sufficiently accurate results for most combinations of
parameters.\cite{BLG2020}

In this paper, we present the results of our investigation of the interaction
of light, beyond the
WFA, with one or two infinite arrays of atoms. We focus on two
aspects of this system. The first is to develop a method that accurately
represents the physics with only a finite number of equations. This is not
a trivial undertaking because there are an infinite number of density
matrix elements for more than 1 excitation. Even weak light leads to
an infinite number of excitations although only a finite fraction of the
atoms are excited.
To model this system,
we extended the higher order mean-field theory in Ref.~\cite{RS12021} to use a mixed order method for the infinite array. The main idea
is to have pair-wise expectation values $\langle \hat{A_n}\hat{B_m}\rangle$
for atoms $n$ and $m$ smoothly transition to products
$\langle \hat{A_n}\rangle\langle\hat{B_m}\rangle$ as the atom separation
increases. The second is to explore the trends for the coherent
reflection or transmission from the array(s) as well as the incoherent
scattering. In this, we somewhat overlap with previous results in
Ref.~\cite{BLG2020} for a single array.

All of the cases treated here
have the atoms perfectly placed on an infinite lattice and the recoil
of the atoms is neglected.
For simplicity, all examples are for a square lattice with the plane wave light
incident perpendicular to the arrays. The details of some phenomena
depend on these specific conditions (for example, the reflection or transmission
of light not normally incident on the array) but there does not appear
to be qualitative changes. Although perfect, uniformly illuminated,
infinite arrays are
not experimentally accessible,
the simplifications that result from this
ideal case could aid in interpreting results from finite, imperfect
arrays. For 1 array, we explored the coherent reflection and transmission
and the incoherent scattering of the photons as a function of detuning
and intensity.
As an example, we showed how the 100\% coherent
reflection on resonance in the WFA changes to incoherent scattering and
some transmission as the intensity increases.
As another example, we found that the lowest-order
mean field approximation
overestimates the incoherent scattering probability by 15\% in the
limit of very low intensity. For 2 arrays, we found that the effects beyond the
WFA can be present at very small incident intensity.

\section{Basic Theory}\label{sec:BT}

In all that follows, we will assume the atoms couple to the light through
a closed two level transition. Also, the atoms will be driven by a classical
plane wave normally incident on the atom arrays to simplify some of the
derivations.
For the $n$-th atom, the ground and excited states are $|g_n\rangle$ and
$|e_n\rangle$. The operators used below follow the definition
\begin{equation}
\hat{e}_n\equiv  |e_n\rangle\langle e_n|\qquad
\hat{\sigma}^-_n\equiv |g_n\rangle\langle e_n|\qquad
\hat{\sigma}^+_n\equiv|e_n\rangle\langle g_n|.
\end{equation}
The position of the $n$th atom is $\bm{R}_n$.

The basic theory is identical to that in Ref.~\cite{RS12021} where the
density matrix equations are converted into equations of motion of the
expectation values of operators. It also uses cumulants, Ref.~\cite{RK11962},
to reduce expectation values of products of operators into products of
lower order expectation values
as done in
Refs.~\cite{FY11999,CYL2004,ROY2023,ROYb2023,LY12012,KR12015,
KK12018,KRK2019,QR12019,OMG2019,HPR2020,SSF2020}.
We will call the replacement of pairwise
expectation values by the product of the expectation value
(e.g. $\langle \hat{A}\hat{B}\rangle\to \langle \hat{A}\rangle\langle
\hat{B}\rangle$) the mean field or mean field-1 (MF1) approximation.
We will call the replacement of triple expectation values by products
of pairwise and single expectation values the mean field-2 (MF2)
approximation
(e.g. $\langle \hat{A}\hat{B}\hat{C}\rangle \to
\langle \hat{A}\hat{B}\rangle\langle\hat{C}\rangle 
+ \langle \hat{A}\rangle\langle\hat{B}\hat{C}\rangle
+ \langle \hat{B}\rangle\langle\hat{A}\hat{C}\rangle 
-2\langle \hat{A}\rangle\langle\hat{B}\rangle\langle\hat{C}\rangle$).
The weak field approximation (WFA) obtains when
all operator products and all $\langle \hat{e}\rangle$ are set to zero.

The coupling between atom pairs through the quantized electromagnetic field
is through the dyadic Green's function. For $m\neq n$, we will define
\begin{eqnarray}
g_{mn}&=&g(\bm{R}_m-\bm{R}_n)\label{Eqgdef1}\\
g(\bm{R}) &=&\frac{\Gamma}{2}\left[h_0^{(1)}(s)+
\frac{3\hat{R}\cdot\hat{d}^*\hat{R}\cdot\hat{d}-1}{2}
h_2^{(1)}(s)\right]\label{Eqgdef2}
\end{eqnarray}
with $\hat{d}$ the dipole unit vector,
$s=kR$, $\hat{R}=\bm{R}/R$, and the $h_\ell^{(1)}(s)$ the outgoing spherical
Hankel function of angular momentum $\ell$:
$h_0^{(1)}(s)=e^{is}/[is]$ and $h_2^{(1)}(s) = (-3i/s^3 - 3/s^2 + i/s)e^{is}$. 
The $g(\bm{R})$ is proportional to the propagator that gives the
electric field at $\bm{R}$ given a dipole at the origin\cite{JDJ1999}.
For
a $\Delta M =0$ transition, $\hat{d}=\hat{z}$ and the coefficient
of the $h_2^{(1)}$ Bessel function is $P_2(\cos (\theta ))=
(3\cos^2(\theta )-1)/2$ where $\cos(\theta )=Z/R$. For
a $\Delta M =\pm 1$ transition, the coefficient
of the $h_2^{(1)}$ Bessel function is $-(1/2)P_2(\cos (\theta ))=
(1-3\cos^2(\theta ))/4$.

For simplicity, the atoms will be on a square array with separation
$a$. We will have the atoms at the position
$\bm{R}_n=(x_0,n_ya,n_za)$ where $x_0=0$ for the single array
calculations and $x_0=0$ or $L$ for the two array calculations.

Some simplifications to one- and two-atom expectation values occur
due to Bloch's theorem
when the light is normally incident on the array. The first is that
the one-atom expectation values are independent of position in that
array. For example, for the single array calculations, we will use:
\begin{equation}
\langle\hat{e}_n\rangle = \langle\hat{e}_0\rangle\equiv
\langle\hat{e}\rangle \qquad {\rm and}
\qquad \langle\hat{\sigma}^\pm_n\rangle = \langle\hat{\sigma}^\pm_0\rangle
\equiv\langle\hat{\sigma}^\pm\rangle.
\label{EqOne0}\end{equation}
The other is that two-atom expectation values can be shifted so that one
operator is at the origin for that array.
For example, for the single array calculations, we will use
\begin{equation}
\langle\hat{A}_{n_y,n_z}\hat{B}_{m_y,m_z}\rangle =
\langle\hat{A}_0\hat{B}_{m_y-n_y,m_z-n_z}\rangle\label{EqTwo0}
\end{equation}
or the index $0$ for the $\hat{B}$ and the index $n_y-m_y,n_z-m_z$ 
for the $\hat{A}$.

The equations of motion for single atom operators ignoring the
interaction between atoms are
\begin{eqnarray}
\frac{d\langle\hat{\sigma}^-\rangle}{dt}&=&[i\Delta-\frac{\Gamma}{2}]
\langle\hat{\sigma}^-\rangle+
i\frac{\Omega}{2}
(2\langle\hat{e}\rangle -1)\nonumber\\
\frac{d\langle\hat{e}\rangle}{dt}&=&-\Gamma\langle\hat{e}\rangle +
i\frac{\Omega^*}{2} \langle\hat{\sigma}^-\rangle-
i\frac{\Omega}{2}\langle\hat{\sigma}^+\rangle
\label{eq:1op}
\end{eqnarray}
where the equation for $\langle\hat{\sigma}^+\rangle$ is the complex conjugate
of the first. When finding the equations for products of operators,
use these equations with the product rule for derivatives.

\section{MF1 equations of motion}

In the time dependent equations for the $\langle\hat{e}\rangle$
and $\langle\hat{\sigma}^\pm\rangle$ expectation
values, only the $0$ index terms contribute as discussed above.
The MF1 approximation leads to the replacement of pairwise expectation
values: $\langle \hat{A}_n\hat{B}_m\rangle\to \langle \hat{A}_n
\rangle\langle\hat{B}_m\rangle\to \langle \hat{A}_0\rangle\langle\hat{B}_0
\rangle \equiv \langle\hat{A}\rangle\langle
\hat{B}\rangle$, see Eq.~(\ref{EqOne0}).

\subsection{One array}

The differential equations for the one atom expectation values have one
atom terms from the interaction with the laser and the single photon
decays in the Lindbladian as well as two atom terms, Eq.~(\ref{eq:App21}),
from the dipole-dipole interactions.
The two atom terms in Eq.~(\ref{eq:App21}) become independent of $m$
for MF1 and the sum over all atoms $m\neq 0$ leads to the definition:
\begin{equation}\label{eq:scrg}
{\cal G}\equiv\sum_{m\neq 0}g(\bm{R}_m)\equiv i\delta +\frac{\gamma}{2}.
\end{equation}
where Ref.~\cite{SWL2017} showed
$\gamma/\Gamma = (3/[4\pi ])(\lambda /a)^2-1$.
We discuss the numerical evaluation of ${\cal G}$ in Appendix~\ref{sec:numapp}.
The equations of motion are
\begin{eqnarray}
\frac{d\langle\hat{\sigma}^-\rangle}{dt}&=&\left[ i\Delta-\frac{\Gamma}{2}\right]
\langle\hat{\sigma}^-\rangle+
i\frac{\bar{\Omega}}{2}
(2\langle\hat{e}\rangle -1)\nonumber\\
\frac{d\langle\hat{e}\rangle}{dt}&=&-\Gamma\langle\hat{e}\rangle +
i\frac{\bar{\Omega}^*}{2} \langle\hat{\sigma}^-\rangle-
i\frac{\bar{\Omega}}{2}\langle\hat{\sigma}^+\rangle
\label{EqEOM1}
\end{eqnarray}
where $\bar{\Omega}=\Omega - 2i {\cal G}\langle\hat{\sigma}^-\rangle$
and the equation for $\langle\hat{\sigma}^+\rangle$ is the complex conjugate
of the first. The $\bar{\Omega}$ can be thought of as the total field
from the laser plus that from all other dipoles.
The steady state values can be found by setting the left
hand side equal to zero with the solution being found using a
Newton's iterative method.

The coherent reflection (${\sf R}$), coherent transmission (${\sf T}$),
and incoherent scattering (${\sf S}$)
probability of light can be determined from derivations in
Appendix~\ref{sec:QLI}.
For the coherent reflection and transmission probabilities,
Eq.~(\ref{eq:cohsc}) gives the far-field form of the electric field
from which
\begin{eqnarray}
{\sf R} = |{\cal R}|^2\qquad &{\rm and}&\qquad {\sf T} = |1+{\cal R}|^2\\
{\cal R}&=&-\frac{3i\Gamma\pi}{\Omega k^2 a^2}\langle\hat{\sigma}^-\rangle\label{eq:ref1}
\end{eqnarray}
with ${\cal R}$ derived in Eq.~(\ref{eq:cohr}). The incoherent scattering
is given by Eq.~(\ref{eq:incsc}) because ${\cal Q}^{(2)}=0$ at the
MF1 level:
\begin{equation}
{\sf S} = 2\times 3\pi\left( \frac{\Gamma}{\Omega k a}\right)^2
[\langle \hat{e}\rangle
-|\langle\hat{\sigma}^+\rangle |^2]
\end{equation}
where the factor of 2 accounts for scattering in both the $\pm x$ directions.

\subsection{Two arrays}

For this case, we will designate the array at $x=0$ to be $\alpha$ and
that at $x=L$ to be $\beta$. The expectation values of the
operators in each array can be different
so we need to distinguish the array with a subscript, for example,
$\langle\hat{e}_\alpha\rangle$. There is a new constant due to the interaction
between the two arrays:
\begin{equation}
{\cal \bar{G}}\equiv e^{-ikL}\sum_{m_\beta}g(\bm{R}_{m_\beta})
\end{equation}
where the $m_\beta$ means to sum over all of the atoms in the array at $x=L$
and the $\bm{R}_{m_\beta}$ are the positions measured relative to the
$0$ atom in array $\alpha$. We have defined ${\cal \bar{G}}$ with a factor
of $\exp (-ikL)$ so that the main dependence on array separation is visible
in the equations of motion.
We discuss the numerical evaluation of
${\cal \bar{G}}$ in Appendix~\ref{sec:numapp}.
If $L$ is larger than a few $\lambda$, the sum is,
to an excellent approximation, ${\cal \bar{G}}=(3\pi /2)\Gamma /(k a)^2$,
see Appendix~\ref{sec:numapp}.
The equations of motion are
\begin{eqnarray}
\frac{d\langle\hat{\sigma}_\alpha^-\rangle}{dt}&=&\left[ i\Delta-\frac{\Gamma}{2}\right]
\langle\hat{\sigma}_\alpha^-\rangle+
i\frac{\bar{\Omega}_\alpha}{2}
(2\langle\hat{e}_\alpha\rangle -1)\nonumber\\
\frac{d\langle\hat{e}_\alpha\rangle}{dt}&=&-\Gamma\langle\hat{e}_\alpha\rangle +
i\frac{\bar{\Omega}_\alpha^*}{2} \langle\hat{\sigma}_\alpha^-\rangle-
i\frac{\bar{\Omega}_\alpha}{2}\langle\hat{\sigma}_\alpha^+\rangle\nonumber\\
\frac{d\langle\hat{\sigma}_\beta^-\rangle}{dt}&=&\left[ i\Delta-\frac{\Gamma}{2}\right]
\langle\hat{\sigma}_\beta^-\rangle+
i\frac{\bar{\Omega}_\beta}{2}
(2\langle\hat{e}_\beta\rangle -1)\nonumber\\
\frac{d\langle\hat{e}_\beta\rangle}{dt}&=&-\Gamma\langle\hat{e}_\beta\rangle +
i\frac{\bar{\Omega}_\beta^*}{2} \langle\hat{\sigma}_\beta^-\rangle-
i\frac{\bar{\Omega}_\beta}{2}\langle\hat{\sigma}_\beta^+\rangle
\label{EqEOM2}
\end{eqnarray}
where the effective fields at array $\alpha$ is $\bar{\Omega}_\alpha = \Omega-2i({\cal G}\langle\hat{\sigma}_\alpha^-
\rangle +\bar{\cal G}e^{ikL}\langle\hat{\sigma}_\beta^-\rangle )$ and at
array $\beta$ is
$\bar{\Omega}_\beta = \Omega e^{ikL}-2i({\cal G}\langle\hat{\sigma}_\beta^-
\rangle +\bar{\cal G}e^{ikL}\langle\hat{\sigma}_\alpha^-\rangle )$.
The coupling between the two arrays is through the $\bar{\Omega}$ parameters.

As with the previous section, the coherent reflection, coherent transmission,
and incoherent scattering probability can be determined from derivations in
Appendix~\ref{sec:QLI}.
For the coherent reflection and transmission probabilities,
Eq.~(\ref{eq:cohsc2}) gives the far-field form of the electric field
from which
\begin{eqnarray}
{\sf R} &=& |{\cal R}_\alpha +{\cal R}_\beta e^{ikL}|^2\\
{\sf T} &=& |1+{\cal R}_\alpha +{\cal R}_\beta e^{-ikL}|^2
\end{eqnarray}
with ${\cal R}$ given in Eq.~(\ref{eq:ref1}) with the expectation
value of the approriate array. As with the previous section,
the incoherent scattering
is given by Eq.~(\ref{eq:incsc}) because ${\cal Q}^{(2)}=0$ at the
MF1 level:
\begin{equation}
{\sf S} = 2\times 3\pi\left( \frac{\Gamma}{\Omega k a}\right)^2
[\langle \hat{e}_\alpha\rangle
-|\langle\hat{\sigma}_\alpha^+\rangle |^2
+\langle \hat{e}_\beta\rangle
-|\langle\hat{\sigma}_\beta^+\rangle |^2]
\end{equation}
where the factor of 2 accounts for scattering in both the $\pm x$ directions.

An interesting parameter is the buildup of coherent light between the arrays.
When the position is not too close to the arrays, the coherent electric field
between the arrays is
\begin{equation}
\vec{E}=\hat{\varepsilon}\left[ (1+{\cal R}_\alpha )e^{ikx}+{\cal R}_\beta
e^{ikL}e^{-ikx}\right]
\end{equation}
where $\hat{\varepsilon}$ is the polarization. The scaled, average intensity
between the arrays can be found by taking the magnitude squared and averaging
between 0 and $L$ giving
\begin{equation}
\langle I\rangle /I_{inc} = |1+{\cal R}_\alpha |^2 + |{\cal R}_\beta |^2
\end{equation}
when the distance between the arrays is nearly an integer number of wavelengths.

\section{MF2 equations of motion}

In the MF2 approximation, the differential equations for the single atom
expectation values do not directly contain an approximation. The
differential equations for two atom expectation values have terms that
arise from the one
atom terms in the Hamiltonian (from the laser) and the Lindbladian
(from the single atom decays) which, because they only contain
pair expectation values, do not have approximations. The terms that
arise from the dipole-dipole interactions have two and three atom
expectation values. The three atom terms are approximated as
two and one atom expectation values as discussed in Appendix~\ref{sec:twoat}.

The only important two atom terms arising from the dipole-dipole interactions are:
\begin{eqnarray}
\frac{d\langle\hat{e}_0\hat{\sigma}^-_n\rangle}{dt}&=&-g^*_{n0}
\langle\hat{e}_n\hat{\sigma}^-_0\rangle-V_1-V_2+2V_3-V_4\\
\frac{d\langle\hat{\sigma}^-_0\hat{\sigma}^-_n\rangle}{dt}&=&2V_5-V_6+2V_7-V_8\\
\frac{d\langle\hat{e}_0\hat{e}_n\rangle}{dt}&=&-V_9-V_{10}-V_{11}-V_{12}\\
\frac{d\langle\hat{\sigma}^-_0\hat{\sigma}^+_n\rangle}{dt}&=&
(g_{n0}+g^*_{n0})(2\langle\hat{e}_0\hat{e}_n\rangle-\langle\hat{e}_0\rangle )
\nonumber\\
&\null &+2V_{13}-V_{14}+2V_{15}-V_{16}
\end{eqnarray}
with the definitions and derivations in Appendix~\ref{sec:twoat}. All other
expectation values can be derived from these.
Including the equations of motion from the single atom operators, use
Eqs.~(\ref{eq:1op}) with the product rule for derivatives. For example,
the contribution to the equation for 
$\langle\hat{e}_0\hat{e}_n\rangle$ is
\begin{eqnarray}
\frac{d\langle\hat{e}_0\hat{e}_n\rangle}{dt}=-2\Gamma \langle\hat{e}_0\hat{e}_n\rangle &+&i\frac{\Omega^*}{2}(\langle \hat{\sigma}^-_0\hat{e}_n\rangle
+ \langle\hat{e}_0\hat{\sigma} ^-_n\rangle )\nonumber\\
&-&i\frac{\Omega}{2}(\langle \hat{\sigma}^+_0\hat{e}_n\rangle
+ \langle\hat{e}_0\hat{\sigma} ^+_n\rangle )
\end{eqnarray}

For the MF2 equations, we have not derived the equations for two arrays.
The results for one array changed by at most 10-20\% going from MF1 to
MF2 for some values
of detuning and Rabi frequency. Going from 1 to 2 arrays will lead to
twice as many single atom expectation values and $4\times$ as many
pair expectation values. In addition, there will be at least $8\times$
the required CPU time. We estimated that the computer requirements
were somewhat beyond our local resources and, therefore, did not attempt
an MF2 calculation for 2 arrays.

\section{Results}\label{sec:Re}

In this section, we present the results of calculations for light incident
on one or two atom arrays. We are specifically interested in how the
light intensity changes the results from that in the weak field limit.

\subsection{One Infinite Array: ${\sf R}$, ${\sf T}$, ${\sf S}$}\label{sec:onearr}

In this section, we consider the coherent reflection (${\sf R}$), coherent
transmission (${\sf T}$), and incoherent scattering (${\sf S}$) of
plane wave light normally incident on a single
layer atom array. The atoms will be on a square array with separation
$a<\lambda$ and approximated as a two-level system.
Reference~\cite{BGA2016} predicted the total reflection of light for the
correct detuning from a
perfect, single layer atomic array and this system was
further explored in Ref.~\cite{SWL2017}. Substantial reflection was experimentally
measured from a single layer atom array.\cite{RWR2020}
The calculations\cite{BGA2016,SWL2017} were in the weak
laser limit where multiple atom excitations is ignored. A key question is
how well does the weak laser approximation work as a function of intensity and how
does the weak laser approximation fail with increasing intensity.
Reference~\cite{BLG2020} presented some results for a Gaussian beam
incident on a $10\times 10$ array for incident intensity up to
$100\; I_{sat}$ (see their Fig.~6). They addressed similar physics questions
to those in this section using (what we call) the MF1 approximation and
found similar results.

\begin{figure}
\resizebox{80mm}{!}{\includegraphics{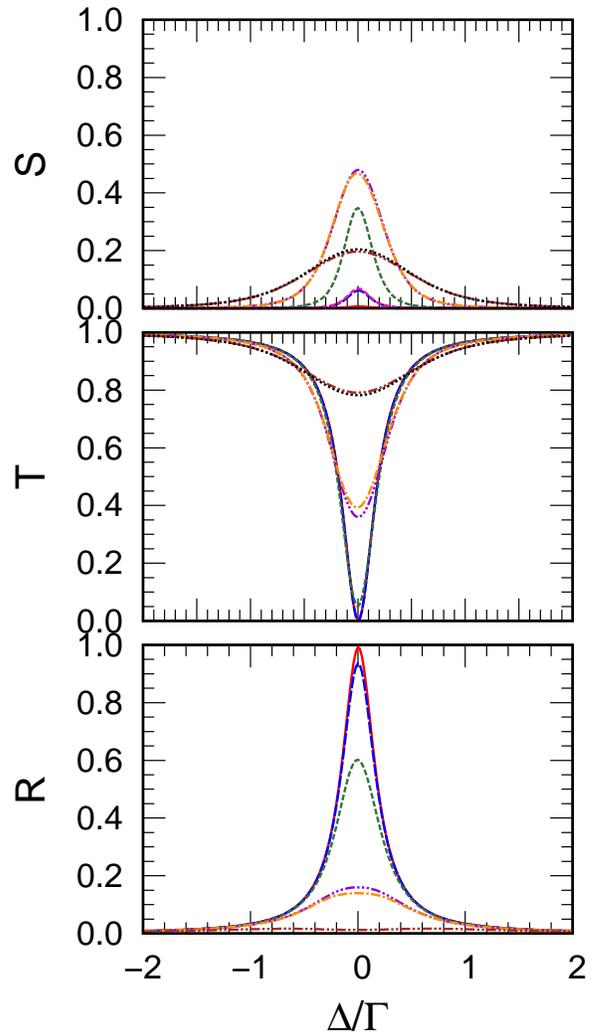}}
\caption{\label{fig:1arr}
The coherently reflected (${\sf R}$), coherently
transmitted (${\sf T}$), and incoherently
scattered (${\sf S}$) probabilities
versus detuning, $\Delta$, for different intensities and approximation
level. The intensity is stepped by powers of 10 and given in terms of
the saturation intensity: $I/I_{sat}=2\Omega^2/\Gamma^2 $.
The MF2 calculations are red solid
($2\times 10^{-4}$), blue long dash ($2\times 10^{-3}$),
green dash (0.02),
orange dash-dot (0.2), and maroon dash-dot-dot (2). In all plots,
the purple long dash is for MF1 with intensity 0.2. In the 
${\sf T}$ and ${\sf S}$ plots, the black dots is for MF1 with intensity 2.
In the ${\sf S}$ plot, the pink dash is for MF1 with intensity
$2\times 10^{-3}$.
}
\end{figure}

It seems clear that the amount of reflection will degrade as the intensity
increases. At low intensity, the atoms can be approximated as harmonic
oscillators which is assumed in Refs.~\cite{BGA2016,SWL2017}. The two-state
character of the atoms become increasingly important as the Rabi frequency
approaches $\Gamma$.
To investigate the role of intensity, we performed MF1 and MF2
calculations of light on a perfect, square array with separation $a=0.8\;\lambda$.
Calculations with separation $0.4$ and $0.6\;\lambda$ gave similar results.
In both MF1 and MF2 calculations, we solved the time dependent equations until the
solutions reached a steady state. The calculations were done with laser
intensities $2\times 10^{-8}$, $2\times 10^{-7}$, $2\times 10^{-6}$,
..., 20, 200 $I_{sat}$. The Rabi frequency
is $\Omega = \Gamma \sqrt{0.5I/I_{sat}}$ giving $10^{-4}$, $\sqrt{10}\times 
10^{-4}$,
$10^{-3}$, ..., $\sqrt{10}$, 10 $\Gamma$.

Figure~\ref{fig:1arr}
shows the transmitted, reflected, and scattered
probabilities versus detuning for different intensities. We only show
the MF1 calculations where there is a clear difference from that for MF2.
Even when there is a clear difference, the difference is not large.
On resonance, there is a larger proportion of photons transmitted or scattered
and smaller proportion
reflected as the laser intensity increases. This trend is not surprising. Even for these
somewhat larger intensities, up to $I=0.2I_{sat}$,
the effect of intensity
is larger for reflection than transmission; for example, the
green dash curve (MF2, $I=0.02I_{sat}$, $\Omega = 0.1\Gamma$)
has approximately 34\% of the photons
scattered, 5\% transmitted, and 61\% reflected. That is, there is still only a 
few coherently transmitted photons, so most of the photons not reflected
are scattered. This figure indicates that the laser intensity needs to be
less than $\sim 0.002\; I_{sat}$, $\Omega\sim 0.03\; \Gamma$,
to keep the fraction of scattered photons
less than $\sim 10$\%. These values depend on the spacing of the lattice.
For $a=0.6\; \lambda$, there is $\sim 5\times $ less scattering for low
intensity but becomes similar at higher intensity. For example, at
$0.002I_{sat}$, there is $\simeq 6$\% ($\simeq 1.2$\%) scattering
for $0.8\lambda$ ($0.6\lambda$) while at $0.2I_{sat}$ both
have a peak scattering of $\sim 40$\%.

As the laser intensity increases from small values, the fraction of incoherently
scattered photons is proportional to the laser intensity times the square of the
reflection probability. Thus, the {\it fraction} of photons that
are incoherently scattered goes to zero as the laser intensity is decreased.
For weak laser intensity, there is little change
to the transmitted probability; the largest change as the intensity increases is a
decrease in the coherent reflected probability.
For example, at $\Delta =0$ for MF2, there is 6.2\% incoherently scattered photons,
93.7\% reflected, and 0.1\% transmitted for $I=0.002\; I_{sat}$ (i.e.
$\Omega\simeq 0.032\;\Gamma $) while
there was 0.67\% scattered and 99.3\% reflected for 10$\times$ smaller
intensity (i.e. $\Omega = 0.01\;\Gamma $).
It is interesting that the MF1 and MF2 incoherently scattered probability
has a ratio at the peak that is independent of the laser intensity for
small intensity: ${\sf S}(MF1)/{\sf S}(MF2)=1.15$
as $I\to 0$. This is a surprising
result because one might expect that MF1 becomes a better approximation
as the fraction of atoms excited is decreased. While this is correct for
reflection and transmission probabilities,
it is not correct for incoherent scattering because the scattered
probability depends on two atom correlations.

As expected, the resonance width increases as the intensity increases.
For smaller intensity,
this is more due to increased decoherence due to incoherent scattering
than to power broadening. For larger intensity, power broadening
becomes increasingly relevant.

Interestingly, as the intensity continues to increase, the
fraction of photons on resonance that are scattered starts to decrease.
This is mainly due to the rate of incoming photons becoming larger
than the rate they can be incoherently scattered.
Also, as expected, the fraction of photons reflected becomes quite small while
the fraction transmitted increases toward 1 as the intensity increases.

\subsection{Two Parallel Arrays}\label{sec:2par}

Two parallel arrays can act as a cavity. This can lead to large intensity
between the arrays which can affect the reflection, transmission, and
incoherent scattering properties at much smaller incident intensities.
As an example, Ref.~\cite{JR12019} investigated several aspects of weak
light interacting with
two-dimensional atomic lattices starting with a pair of lattices and
going to stacks of lattices.
All of the calculations in this section use MF1 with the arrays separated
by $5.01\;\lambda $ and the atom separation within an array $a=0.8\;\lambda $.
The cavity enhancement of intensity between the array and the narrowness of
the resonance depends on how close the separation is to an integer
(or half-integer) number of wavelengths; the value $5.01\;\lambda $ was
chosen to give a cavity enhancement of a few 100 on resonance.
From the previous section, there is approximately 15\% more incoherently scattered light
when using the MF1 approximation compared to MF2 for one array. Therefore,
the intensity effects will be somewhat exaggerated in these calculations but
the approximate sizes and the trends should be correct.

We first derive the expected result in the weak field approximation when
the array separation is much larger than the atom spacing within one array,
$L\gg a$. In this limit, the scattering from each array can be done self
consistently using the notation of Appendix~\ref{sec:QLI}:
\begin{eqnarray}
{\cal R}&=&-\frac{3i\Gamma\pi}{\Omega k^2 a^2}
\frac{\Omega /2}{\Delta+i(\Gamma /2) +i{\cal G}}\nonumber\\
\langle\eta^+_E\rangle(x<L)&=&\hat{\varepsilon }(e^{ikx}+{\cal R}e^{ik|x|}
+A_\alpha [e^{-ikx}+{\cal R}e^{ik|x|}])\nonumber\\
\langle\eta^+_E\rangle(x>0)&=&\hat{\varepsilon }A_\beta (e^{ik(x-L)}+{\cal R}e^{ik|x-L|})
\end{eqnarray}
where we used the weak field approximation to calculate ${\cal R}$ and
used the fact that the reflection amplitude is the same for both left and right
going waves. The two coefficients, $A_{\alpha ,\beta}$, are determined by making
the two expression equal in the region $0<x<L$.
This leads to the two equations
$1+{\cal R}+{\cal R}A_\alpha =A_\beta e^{-ikL}$ and $A_\alpha =A_\beta {\cal R}
e^{ikL}$ giving
\begin{eqnarray}
A_\alpha &=& \frac{{\cal R} (1+{\cal R})}{1-{\cal R}^2e^{2ikL}}e^{2ikL}\\
A_\beta &=& \frac{1+{\cal R}}{1-{\cal R}^2e^{2ikL}}e^{ikL}
\end{eqnarray}
The total reflection and transmission
amplitudes are:
\begin{eqnarray}
{\cal R}_{tot} &=& {\cal R} + (1+{\cal R})A_\alpha\nonumber\\
{\cal T}_{tot}&=& (1+{\cal R})A_\beta e^{-ikL}
\end{eqnarray}
The relative average of the intensity between the arrays (average intensity between
the arrays divided by incident intensity) can be found using
$\langle \eta_E^+\rangle (x>0)$ by taking the magnitude squared and averaging
between 0 to $L$ giving
\begin{equation}
\langle I\rangle /I_{inc} = |A_\beta|^2 (1 + |{\cal R}|^2)
\end{equation}
in the weak field limit.

\begin{figure}
\resizebox{80mm}{!}{\includegraphics{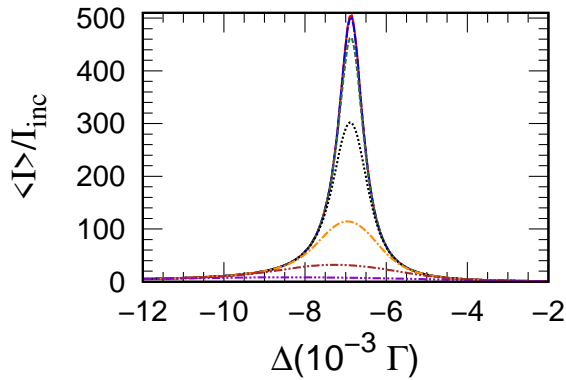}}
\caption{\label{fig:2arrbet}
The relative average intensity between the arrays versus detuning for several
incident intensities (in terms of $I_{sat}$): red solid (weak field approximation),
blue dashed ($2\times 10^{-9}$), green short dash ($2\times 10^{-8}$),
black dotted ($2\times 10^{-7}$), orange dash-dot ($2\times 10^{-6}$),
maroon dash-dot-dot ($2\times 10^{-5}$), and purple dash-dot-dot ($2\times 10^{-4}$).
The arrays are separated by $5.01\;\lambda$ and the atom separation
$a=0.8\;\lambda $.
}
\end{figure}

\begin{figure}
\resizebox{80mm}{!}{\includegraphics{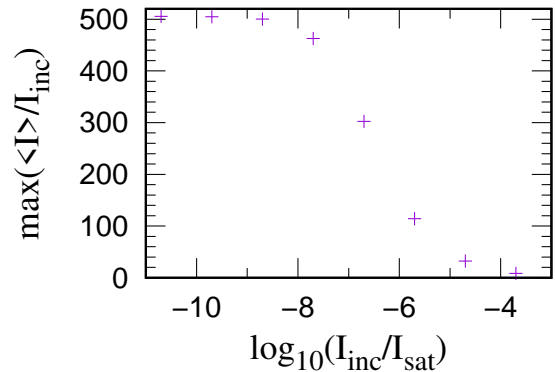}}
\caption{\label{fig:2arrbeti}
The peak relative average intensity between the arrays versus the
incident intensity.
}
\end{figure}

As with all cavities, the largest response is when the reflection probability is close
to 1 which implies array separations close to an integer number of wave
lengths. This will lead to a transmission probability with a narrow resonance and
a sharp peak in the average intensity between the arrays. The increased intensity
between the arrays leads to a
conflict for the response of the array pair.
The peak intensity between the arrays increases
with the narrowness of the line. But as the incident intensity,
$I_{inc}$, increases, there will be
more incoherent scattering from the effects discussed in Sec.~\ref{sec:onearr}
which will lead to a broadening of the line and a decrease in relative intensity
between the arrays.

This can be seen in Fig.~\ref{fig:2arrbet} where the average intensity between
the arrays divided by the incident intensity is plotted versus detuning for
the weak field approximation and
several incident intensities, from $2\times 10^{-9}$ to $2\times 10^{-4}\; I_{sat}$.
The results approach that for the weak field approximation as the incident
intensity decreases.
As expected, the relative
intensity between the arrays decreases with increasing incident
intensity.
The surprising aspect might be at how small an intensity the decrease becomes noticeable.
At $2\times 10^{-8}\; I_{sat}$ incident intensity, the peak value is decreased by
approximately 8.3\% from the weak field limit. This is approximately the same factor
of incoherently scattered photons. In fact, for low intensities, the relative
difference in the intensity from the weak field approximation approximately equals
the fraction of photons scattered for detunings in the neighborhood of the
resonance. From Sec.~\ref{sec:onearr},
approximately 10\% scattering occurred for $2\times 10^{-3}\; I_{sat}$. Making a
crude estimate that the intensity is $500\times $ larger and there are 2 arrays, one
might expect an effective intensity $1000\times $ larger than
$2\times 10^{-8}\; I_{sat}$ giving an effective
intensity for incoherent scattering of $2\times 10^{-5}\; I_{sat}$. 
For the two arrays, we expected an incoherently scattered probability closer to 0.083\%.
Thus, the incoherent scattering for two arrays could be much larger than from simple
estimates.

Figure~\ref{fig:2arrbeti} shows the peak of the relative average intensity
between the arrays plotted versus the incident intensity for intensities
from $2\times 10^{-11}$ to $2\times 10^{-4}\; I_{sat}$. At low intensities, it goes
to the value for the weak field approximation, $\sim 500$.
At $2\times 10^{-4}\; I_{sat}$, it has decreased by a factor of $\sim 60$
to $\sim 8$ which is still a large enhancement.

Changing the
array separation to $5.01414\; \lambda $ (i.e. increase the displacement from integer
wavelength by a factor of $\sqrt{2}$) shifts the position of the peak and 
lowers the maximum value
by a factor of 2. For $2\times 10^{-8}\; I_{sat}$, the peak fraction of
scattered photons was decreased to 2.4\%, i.e by a factor of approximately 3.5.
Changing the array separation
to $5.02\; \lambda $ shifts the position of the peak and lowers the maximum value
by a factor of 4. For $2\times 10^{-8}\; I_{sat}$, the peak fraction of
scattered photons decreased to 0.61\%, i.e.
by another factor of approximately 3.8.

\begin{figure}
\resizebox{80mm}{!}{\includegraphics{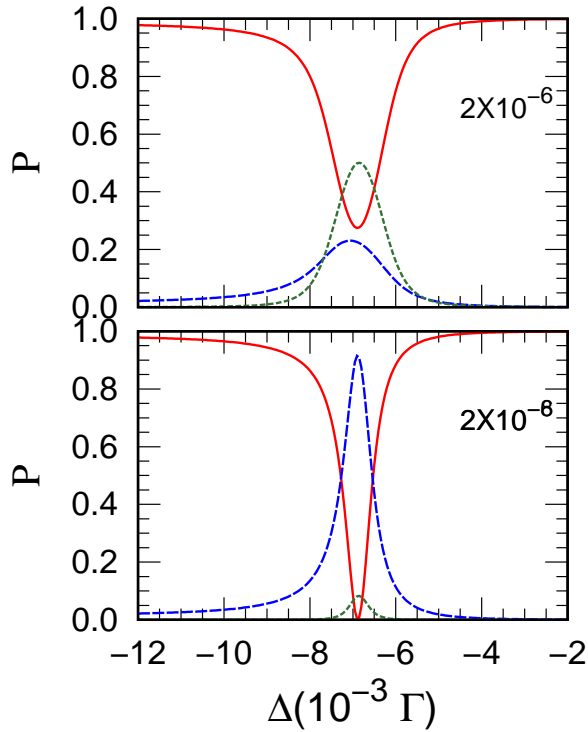}}
\caption{\label{fig:2arrsct}
For a pair of arrays, the reflected (red solid),
transmitted (blue dashed), and incoherently scattered (green short dash)
probabilities versus detuning for an intensity of $2\times 10^{-8}$ and of
$2\times 10^{-6}\; I_{sat}$. The
array properties are the same as for Fig.~\ref{fig:2arrbet}.
}
\end{figure}

Figure~\ref{fig:2arrsct} shows the reflection,
transmission, and incoherent scattering probabilities versus detuning
for two very low intensities. In the weak field limit on resonance,
the reflection probability goes
to 0 and the transmission probability goes to 1.
As with the one array case, the biggest change
at low intensities is the transfer of transmission probability to incoherently
scattered probability. For example, for the $2\times 10^{-8}\; I_{sat}$ case, 
the reflection probability on resonance is less than 1\% while the transmitted
probability has decreased to approximately 92\%. For the $2\times 10^{-6}\; I_{sat}$ case,
the reflection probability on resonance is approximately 28\% which is more than
the approximately 23\% transmitted. Another clear feature is the increase in
resonance width with increasing intensity. Obviously, there is no relevant power
broadening for intensity of $2\times 10^{-6}\; I_{sat}$ but the linewidth is clearly
much larger than that for the weak field limit.

\begin{figure}
\resizebox{80mm}{!}{\includegraphics{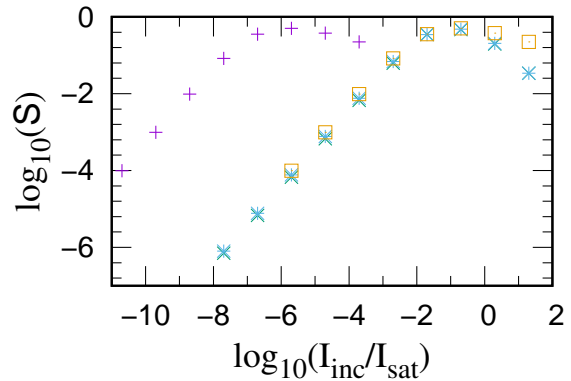}}
\caption{\label{fig:scat_1_2}
The maximum incoherent scattering (${\sf S}$) versus incident intensity for 2
arrays calculated using MF1 (purple $+$) and for 1 array calculated
using MF2 (green $\times$) and MF1 (blue $*$). The orange squares are
the same as the purple $+$ but with $I_{inc}$ multiplied by $10^5$.
}
\end{figure}

Figure~\ref{fig:scat_1_2} shows the maximum incoherent scattering probability
versus the incident intensity for 1 and 2 arrays. This shows the linear
rise of ${\sf S}$ with intensity for small intensities and then the subsequent
decrease as the atom excitation is saturated. It also shows the relatively large
amount of scattering for the 2 array cavity which only has an intensity
enhancement of a factor of $\simeq 500$. The 2 array scattering overlaps the 1
array for small $I_{inc}$ if the 2 array intensity is increased by a factor
of $\sim 10^5$.
Both the average intensity between the arrays and the scattering probabilities show
that even quite weak laser intensity can lead to qualitative changes.

\subsection{Extent of the near field}

In our calculations, the main parameters derive from the asymptotic behavior of the
light. For example, the reflection, transmission, and incoherent scattering
probabilities are all asymptotic properties. Clearly, the light is not
a plane wave in the neighborhood of the atoms. So an important question
is how does the light field become a plane wave as a function of distance
from an atom array. This is a natural question because the near field
could affect results.  For example, we chose the atoms in both arrays to be at the
same $y,z$-positions in Sec.~\ref{sec:2par}. This choice is irrelevant
if the field has become a plane wave over the distance between the
two arrays.

The main dependence
of the decay arises from the periodicity in the light fields due to
the atom array. Because the atom array repeats after a displacement of
$a<\lambda$ in the $y$- or $z$-directions, the light fields must have
$(k_y,k_z)=(2\pi /a)(n_y,n_z)$ with $n_i$ being integers. Only for
$n_y=n_z=0$ is the resulting wave number in the $x$-direction real.
For other values, the $x$-direction will exponentially decrease with
$\kappa_x = k\sqrt{(n_y^2+n^2_z)(\lambda /a )^2-1}$. The slowest
decrease will be when $n_y^2+n_z^2=1$ giving an exponential decrease
\begin{equation}
{\rm near\; field}\sim e^{-\kappa_x x}
\end{equation}
with $\kappa_x=k\sqrt{(\lambda /a )^2-1}$. For the examples in the figures,
we used $a=0.8\lambda$ giving $\kappa_x=4.71/\lambda $. Thus, the size of
the near field has decreased by a factor of $\sim 6\times 10^{-11}$ between
the two arrays in Sec.~\ref{sec:2par}.
We tested this by subtracting the asymptotic form in Eq.~(\ref{eq:cohsc})
from a numerical summation of the first line and exactly found this
exponential decay. This can lead to surprisingly fast decay of the
near field when the atom separation is small. For example, for $a=0.6\lambda$,
$\kappa_x = 8.38/\lambda$ giving a near field decay by a factor of
$\sim 4000$ after only one wavelength.

\subsection{Photon correlations}

These studies were partly motivated by the possibility for investigating
correlations between photons reflected, scattered, or transmitted
through the arrays. There have been several studies for finite
arrays, for example Refs.~\cite{DC12021,PZP2023,WBR2020}.
Unfortunately, the infinite array cases do not
display interesting behavior with regards to, for example, the $g^{(2)}$
functions for photons. One can show that all combinations of
$g^{(2)}$ are equal to 1 for all
times because the range of atom correlations is finite
but the arrays extend to infinity. Thus, only finite size arrays or
focused light beams lead to nontrivial $g^{(2)}$ for the geometry
investigated here.

\section{Summary}

We have developed a computationally tractable theory for light interacting
with infinite atom array(s) beyond the weak field approximation used
in previous investigations. The infinite extent of the arrays provides both
a challenge and a simplification to quantitatively understanding the
dynamics. We mainly focus on the behavior of the coherently reflected
or transmitted photons and the incoherently scattered photons versus
the intensity and detuning of the light with a focus on how larger
intensities modify the transmission, reflection, and incoherent
scattering properties. Results were presented for both
a single atom array as well as a pair of atom arrays with a separation
that leads to cavity enhancement of light between the arrays.

The basic computational tool was the mean field approximation. Calculations
were performed using simple mean field (MF1) as well as a more elaborate
mean field (MF2): in MF1, expectation values of products of operators
are replaced by products of expectation values while MF2 replaces
expectation values of a triple product by pairwise and single
expectation values. One of the main results is the reduction of the
infinite number of equations into a finite number that well represents
this system. We also give practical methods for evaluating the infinite
sums for important parameters.

The basic physical question addressed was how larger intensity modifies the
interaction of the array(s) with light. For a single array,
we showed that the MF1 and MF2
results were nearly the same except for a small range of intensity. As
might be expected, the fraction of incoherently scattered photons increases
with increasing Rabi frequency, $\Omega$, until the intensity is large
enough to begin saturating the transition. At larger intensity, the
coherently reflected and transmitted probabilities are changed in
characteristic ways. For example, on resonance, the coherently
reflected probability decreases from 1 with increasing intensity while
the coherently transmitted probability increases from 0. For the case
of two arrays with separation to give cavity enhancement of photons
between them, the changes to the transmitted and reflected probabilities
are substantial for tiny incident laser intensities.

Data plotted in the figures is available at~\cite{data}.

\begin{acknowledgments}
FR thanks Igor Ferrier-Barbut for a discussion that
motivated the form of Eqs.~(\ref{EqEOM1}) and (\ref{EqEOM2}).
This work was supported by the National Science Foundation under Grant
No. 2109987-PHY.
\end{acknowledgments}

\appendix

\section{Two atom terms}\label{sec:twoat}

This section gives the terms that arise from the dipole-dipole interaction.
In the one atom equations of motion, there are terms that arise from the
interaction between pairs of atoms. Because the single atom expectation
values are independent of atom position, the only important terms are
\begin{eqnarray}
\frac{d\langle\hat{\sigma}^-_0\rangle}{dt}&=&\sum_{m\neq 0}g_{m0}\langle
(2\hat{e}_0-1)\hat{\sigma}^-_m\rangle\nonumber\\
\frac{d\langle\hat{e}_0\rangle}{dt}&=&-\sum_{m\neq 0}
\left[ g_{m0}\langle\hat{\sigma}^+_0\hat{\sigma}^-_m\rangle + g^*_{m0}
\langle\hat{\sigma}^-_0\hat{\sigma}^+_m\rangle\right]\label{eq:App21}
\end{eqnarray}

In the two atom equations of motion,
the dipole-dipole terms in the differential equations are complicated so
we have broken them up into 16 different terms. Only in the first of these,
$V_1$, will we explicitly show the conversion of the expectation value of
triple operators into pairs and singles. Remember all single atom expectation
values can be replaced as $\langle\hat{e}_n\rangle =\langle\hat{e}_0\rangle $.
These terms are
\begin{eqnarray}
V_1&=&\sum_{m\neq n,0}g_{m0}\langle\hat{\sigma}^+_0\hat{\sigma}^-_n
\hat{\sigma}^-_m\rangle\nonumber\\
&\to&\sum_{m\neq n,0}g_{m0}(
\langle\hat{\sigma}^+_0\rangle\langle\hat{\sigma}^-_n\hat{\sigma}^-_m\rangle 
+\langle\hat{\sigma}^+_0\hat{\sigma}^-_n\rangle\langle\hat{\sigma}^-_m\rangle
\nonumber\\ &\null &\hskip 0.2 in
+\langle\hat{\sigma}^+_0\hat{\sigma}^-_m\rangle\langle\hat{\sigma}^-_n\rangle
-2\langle\hat{\sigma}^+_0\rangle\langle\hat{\sigma}^-_n\rangle\langle\hat{\sigma}^-_m\rangle )\nonumber\\
&=&({\cal G}-g_{n0})
(\langle\hat{\sigma}^+_0\hat{\sigma}^-_n\rangle
-2\langle\hat{\sigma}^+_0\rangle\langle\hat{\sigma}^-_0\rangle )\langle
\hat{\sigma}^-_0\rangle \nonumber\\
&\null&+({\cal C}^+_0-g_{n0}\langle\hat{\sigma}^+_0\hat{\sigma}^-_n\rangle 
)\langle\hat{\sigma}^-_0\rangle+{\cal C}_n^{-}\langle\hat{\sigma}^+_0\rangle\\
V_2&=&\sum_{m\neq n,0}g_{m0}^*\langle\hat{\sigma}^-_0\hat{\sigma}^-_n
\hat{\sigma}^+_m\rangle\nonumber\\
&=&({\cal G}^*-g_{n0}^*)
(\langle\hat{\sigma}^-_0\hat{\sigma}^-_n\rangle
-2\langle\hat{\sigma}^-_0\rangle\langle\hat{\sigma}^-_0\rangle )\langle
\hat{\sigma}^+_0\rangle\nonumber\\
&\null&+({\cal C}^{+*}_0-g_{n0}^*\langle\hat{\sigma}^-_0\hat{\sigma}^+_n\rangle
)\langle\hat{\sigma}^-_0\rangle +{\cal C}_n^{+*}\langle\hat{\sigma}^-_0\rangle\\
V_3&=&\sum_{m\neq n,0}g_{mn}\langle\hat{e}_0\hat{e}_n
\hat{\sigma}^-_m\rangle\nonumber\\
&=&({\cal G}-g_{-n0})
(\langle\hat{e}_0\hat{e}_{-n}\rangle
-2\langle\hat{e}_0\rangle\langle\hat{e}_0\rangle )\langle
\hat{\sigma}^-_0\rangle\nonumber\\
&\null&+({\cal C}^0_0-g_{-n0}\langle\hat{e}_0\hat{\sigma}^-_{-n}\rangle
)\langle\hat{e}_0\rangle +{\cal C}_{-n}^{0}\langle\hat{e}_0\rangle\\
V_4&=&\sum_{m\neq n,0}g_{mn}\langle\hat{e}_0\hat{\sigma}^-_m\rangle =
{\cal C}_{-n}^{0}
\end{eqnarray}
\begin{eqnarray}
V_5&=&\sum_{m\neq n,0}g_{m0}\langle\hat{e}_0\hat{\sigma}^-_n
\hat{\sigma}^-_m\rangle\nonumber\\
&=&({\cal G}-g_{n0})
(\langle\hat{e}_0\hat{\sigma}^-_n\rangle
-2\langle\hat{e}_0\rangle\langle\hat{\sigma}^-_0\rangle )\langle
\hat{\sigma}^-_0\rangle\nonumber\\
&\null&+({\cal C}^0_0-g_{n0}\langle\hat{e}_0\hat{\sigma}^-_n\rangle
)\langle\hat{\sigma}^-_0\rangle+{\cal C}_n^{-}\langle\hat{e}_0\rangle\\
V_6&=&\sum_{m\neq n,0}g_{m0}\langle\hat{\sigma}^-_n
\hat{\sigma}^-_m\rangle = {\cal C}_n^-\\
V_7&=&\sum_{m\neq n,0}g_{mn}\langle\hat{\sigma}^-_0\hat{e}_n
\hat{\sigma}^-_m\rangle\nonumber\\
&=&({\cal G}-g_{-n0})
(\langle\hat{e}_0\hat{\sigma}^-_{-n}\rangle
-2\langle\hat{e}_0\rangle\langle\hat{\sigma}^-_0\rangle )\langle
\hat{\sigma}^-_0\rangle\nonumber\\
&\null&+({\cal C}^0_0-g_{-n0}\langle\hat{e}_0\hat{\sigma}^-_{-n}\rangle
)\langle\hat{\sigma}^-_0\rangle+{\cal C}_{-n}^{-}\langle\hat{e}_0\rangle\\
V_8&=&\sum_{m\neq n,0}g_{mn}\langle\hat{\sigma}^-_0
\hat{\sigma}^-_m\rangle= {\cal C}_{-n}^-
\end{eqnarray}
\begin{eqnarray}
V_9&=&\sum_{m\neq n,0}g_{m0}\langle\hat{\sigma}^+_0\hat{e}_n
\hat{\sigma}^-_m\rangle\nonumber\\
&=&({\cal G}-g_{n0})
(\langle\hat{\sigma}^+_0\hat{e}_n\rangle
-2\langle\hat{\sigma}^+_0\rangle\langle\hat{e}_0\rangle )\langle
\hat{\sigma}^-_0\rangle\nonumber\\
&\null&+({\cal C}^+_0-g_{n0}\langle\hat{\sigma}^+_0\hat{\sigma}^-_n\rangle
)\langle\hat{e}_0\rangle+{\cal C}_n^0\langle\hat{\sigma}^+_0\rangle\\
V_{10} &=&\sum_{m\neq n,0}g_{m0}^*\langle\hat{\sigma}^-_0\hat{e}_n
\hat{\sigma}^+_m\rangle = V_9^*\\
V_{11}&=&\sum_{m\neq n,0}g_{mn}\langle\hat{e}_0\hat{\sigma}^+_n
\hat{\sigma}^-_m\rangle\nonumber\\
&=&({\cal G}-g_{-n0})
(\langle\hat{\sigma}^+_0\hat{e}_{-n}\rangle
-2\langle\hat{\sigma}^+_0\rangle\langle\hat{e}_0\rangle )\langle
\hat{\sigma}^-_0\rangle\nonumber\\
&\null&+({\cal C}^+_0-g_{-n0}\langle\hat{\sigma}^+_0\hat{\sigma}^-_{-n}\rangle
)\langle\hat{e}_0\rangle+{\cal C}_{-n}^0\langle\hat{\sigma}^+_0\rangle\\
V_{12} &=& \sum_{m\neq n,0}g_{mn}^*\langle\hat{e}_0\hat{\sigma}^-_n
\hat{\sigma}^+_m\rangle =V_{11}^*
\end{eqnarray}
\begin{eqnarray}
V_{13}&=&\sum_{m\neq n,0}g_{m0}\langle\hat{e}_0\hat{\sigma}^+_n
\hat{\sigma}^-_m\rangle\nonumber\\
&=&({\cal G}-g_{n0})
(\langle\hat{e}_0\hat{\sigma}^+_n\rangle
-2\langle\hat{e}_0\rangle\langle\hat{\sigma}^+_0\rangle )\langle
\hat{\sigma}^-_0\rangle\nonumber\\
&\null&+({\cal C}^0_0-g_{n0}\langle\hat{e}_0\hat{\sigma}^-_n\rangle
)\langle\hat{\sigma}^+_0\rangle+{\cal C}_n^{+}\langle\hat{e}_0\rangle\\
V_{14}&=&\sum_{m\neq n,0}g_{m0}\langle\hat{\sigma}^+_n
\hat{\sigma}^-_m\rangle = {\cal C}_n^+\\
V_{15}&=&\sum_{m\neq n,0}g^*_{mn}\langle\hat{\sigma}^-_0\hat{e}_n
\hat{\sigma}^+_m\rangle\nonumber\\
&=&({\cal G}-g_{-n0})^*
(\langle\hat{e}_0\hat{\sigma}^-_{-n}\rangle
-2\langle\hat{e}_0\rangle\langle\hat{\sigma}^-_0\rangle )\langle
\hat{\sigma}^+_0\rangle\nonumber\\
&\null&+({\cal C}^0_0-g_{-n0}\langle\hat{e}_0\hat{\sigma}^-_{-n}\rangle
)^*\langle\hat{\sigma}^-_0\rangle+{\cal C}_{-n}^{+*}\langle\hat{e}_0\rangle\\
V_{16}&=&\sum_{m\neq n,0}g^*_{mn}\langle\hat{\sigma}^-_0
\hat{\sigma}^+_m\rangle= {\cal C}_{-n}^{+*}
\end{eqnarray}
where the ${\cal G}$ is defined in Eq.~(\ref{eq:scrg}) and subscripts
$-n$ means the atom identified by the point $-n_y,-n_z$.
The other parameters used in these equations are
\begin{eqnarray}
{\cal C}^\pm _n&=&\sum_{m\neq 0,n}g_{m0}\langle
\hat{\sigma}^\pm_n\hat{\sigma}^-_m \rangle\\
{\cal C}^0 _n&=&\sum_{m\neq 0,n}g_{m0}\langle
\hat{e}_n\hat{\sigma}^-_m \rangle
\end{eqnarray}
We discuss the numerical evaluations of these three parameters in
Appendix~\ref{sec:numapp}.

\section{Quantum Light Intensity}\label{sec:QLI}

This section gives the equations for how to calculate the light intensity
in the presence of the atom array(s). The flux of photons will be given
in terms of the expectation values of the atom operators. Some care is
needed to insure the infinite sums are handled correctly. The derivation
is for a plane wave normally incident on the atom array and will compute
the relative photon flux at large distance from the arrays.

Following Ref.~\cite{WR12020}, we start from the field operators
\begin{eqnarray}
\bm{\eta}^+_E(\bm{r})
&=&\hat{\varepsilon}e^{i\bm{k}\cdot\bm{r}}-\frac{2i}{\Omega}
\sum_n\bm{g}_E(\bm{r}_n)\hat{\sigma}^-_n\\
\bm{\eta}^+_B(\bm{r})
&=&\hat{k}\times
\hat{\varepsilon}e^{i\bm{k}\cdot\bm{r}}-\frac{2i}{\Omega}
\sum_n\bm{g}_B(\bm{r}_n)\hat{\sigma}^-_n\\
\bm{g}_E(\bm{R})&=&\frac{3\Gamma}{4}\frac{e^{ikR}}{ikR}[\hat{\varepsilon}-
(\hat{\varepsilon}\cdot\hat{R})\hat{R}]\\
\bm{g}_B(\bm{R})&=&\hat{R}\times\bm{g}_E(\bm{R})
\end{eqnarray}
with $\bm{r}_n=\bm{r}-\bm{R}_n$,
$\hat{\varepsilon}$ being the unit vector for the laser polarization and
the dipole of the atom,
and $r_n\gg\lambda$. Because of the last condition,
the asymptotic, $|\bm{R}|\to\infty$, form
of the electric- and magnetic-field amplitudes, $g_{E,B}(\bm{R})$, are used.
The field operators have been scaled by a factor
proportional to the Rabi frequency so that the coherent reflection,
coherent transmission,
and incoherent scattered probabilities are easily obtained.
For all of the calculations in this paper, the
$\vec{r}=x\hat{k}$ with $x\to\pm\infty$, i.e. the evaluation of the
fields will be at large distance from the array above or below the
origin.
The relative flux of photons in the direction $\hat{k}$ is
\begin{equation}
{\cal F}(\bm{r})
=\frac{1}{2}\hat{k}\cdot\langle\bm{\eta}^-_E\times\bm{\eta}^+_B
-\bm{\eta}^-_B\times\bm{\eta}^+_E\rangle .
\end{equation}
To evaluate the expression, it is simplest to use the classical expression
and the difference from the classical value:
\begin{equation}\label{eq:Fsplt}
{\cal F}(\bm{r})={\cal F}_{cl}(\bm{r}) + {\cal S}(\bm{r})
\end{equation}
with ${\cal F}_{cl}(\bm{r})$ the classical value which is coherent with the
driving laser and ${\cal S}(\bm{r})$ which is incoherent and describes
the scattered photons.

The  classical expression is
\begin{equation}
{\cal F}_{cl}(\bm{r})=\frac{1}{2}\hat{k}\cdot (\langle\bm{\eta}^-_E\rangle
\times\langle\bm{\eta}^+_B\rangle
-\langle\bm{\eta}^-_B\rangle\times\langle\bm{\eta}^+_E\rangle )
\end{equation}
where the large distance expressions can be found analytically. For a single
array, the classical field is
\begin{eqnarray}
\langle\bm{\eta}^+_E\rangle &=&\hat{\varepsilon}e^{ikx}-\frac{2i}{\Omega}
\langle\hat{\sigma}^-\rangle\sum_n\bm{g}_E(\bm{r}_n)\nonumber \\
&\to &\hat{\varepsilon} \left( e^{ikx} + {\cal R}e^{ik|x|}\right)
\label{eq:cohsc}
\end{eqnarray}
with the reflection amplitude, ${\cal R}$, given by
\begin{equation}\label{eq:cohr}
{\cal R}=-\frac{3i\Gamma\pi}{\Omega k^2 a^2}\langle\hat{\sigma}^-\rangle
\end{equation}
because $|x|\gg a$.
This result was found using the relation, $d^2_n=x^2+y_n^2+z_n^2$,
to get
\begin{eqnarray}\label{eq:sumint1}
\sum_n\frac{e^{ikd_n}}{ikd_n}\left(
1 - \frac{z_n^2}{d^2_n}\right)&\to &\frac{1}{ia^2}\int\int
\frac{e^{ikd}}{kd}\left(
1 - \frac{z^2}{d^2}\right) dy dz\nonumber\\
&=&\frac{2\pi}{k^2a^2}e^{i k |x|}
\end{eqnarray}
for $|x|\gg a$. This expression is for linear polarization {\it and} for
circular polarization because replacing $z_n^2$ with $(1/2)(y_n^2+z_n^2)$
leads to the same result.
For a pair of arrays, the classical field is
\begin{equation}\label{eq:cohsc2}
\langle\bm{\eta}^+_E\rangle\to\hat{\varepsilon}\left(
e^{ikx} +{\cal R}_\alpha e^{ik|x|} +{\cal R}_\beta e^{ik|x-L|}\right)
\end{equation}
as $|x|\gg a$ with the subscript on ${\cal R}$ indicating which expectation
value is used in Eq.~(\ref{eq:cohr}).

The scattering terms arise from the cumulant of the
pair-wise expectation values ($\langle\hat{A}\hat{B}\rangle
-\langle\hat{A}\rangle\langle\hat{B}\rangle$) in Eq.~(\ref{eq:Fsplt}):
\begin{equation}
{\cal S}(\bm{r})= {\cal Q}^{(1)}(\bm{r}) + {\cal Q}^{(2)}(\bm{r})
\end{equation}
where the ${\cal Q}^{(1)},{\cal Q}^{(2)}$ are derived separately.
As $|x|\to\infty$, the ${\cal Q}^{(1)},{\cal Q}^{(2)}$ each go to a constant
because there can't be phase information from the incoherent scattering
from all of the atoms in the array.

The one atom terms arise because $\hat{\sigma}^+_n\hat{\sigma}^-_n=\hat{e}_n$
converts a two-operator expectation value to a single operator
expectation value. This term is
\begin{eqnarray}
{\cal Q}^{(1)}&=&\left(\frac{3\Gamma}{2 k\Omega} \right)^2[\langle \hat{e}\rangle
-|\langle\hat{\sigma}^+\rangle |^2]\sum_n(\hat{k}\cdot\hat{r}_n)
\frac{1-|\hat{\varepsilon}\cdot\hat{r}_n |^2}{r_n^2}\nonumber\\
&\to&3\pi\left( \frac{\Gamma}{\Omega k a}\right)^2
[\langle \hat{e}\rangle
-|\langle\hat{\sigma}^+\rangle |^2]\label{eq:incsc}
\end{eqnarray}
where $\vec{r}_n=\bm{r}-\bm{R}_n$ with $x\to\pm \infty$.
This result was found using the relation, $d^2_n=x^2+y_n^2+z_n^2$,
to get
\begin{eqnarray}
\sum_n\frac{|x|}{d_n^3}\left(
1 - \frac{z_n^2}{d^2_n}\right)&\to &\frac{1}{a^2}\int\int
\frac{|x|}{d^3}\left(
1 - \frac{z^2}{d^2}\right) dy dz\nonumber\\
&=&\frac{4\pi}{3a^2}
\end{eqnarray}
in the limit $|x|\to\infty$. This expression is the same for linear
or circular polarization. Note that ${\cal Q}^{(1)}$ is the same
in the $+x$ or $-x$ directions so that the calculation of the total incoherent
scattering probability needs to double these values.

The two atom terms must have the form
\begin{equation}\label{eq:q2mf2}
{\cal Q}^{(2)}\to\sum_{m\neq 0}q_m[\langle\hat{\sigma}^+_0\hat{\sigma}^-_m
\rangle -\langle\hat{\sigma}^+_0\rangle\langle\hat{\sigma}^-_0
\rangle]
\end{equation}
with $q_m$ independent of position. For linear polarization,
\begin{equation}
q_m\to \left(\frac{3\Gamma}{2\Omega k} \right)^2\sum_n
\frac{|x|}{d_n^3}\left(1-\frac{z_n^2}{d_n^2} \right)e^{ik(n_ym_y+n_zm_z)a^2/d_n}
\end{equation}
in the limit $|x|\to\infty $. We have not found a simple closed form solution
for this expression. For example, the $q_m$ can be written in terms of
integrals over $J_0,J_2$ cylindrical Bessel functions or as a series in
powers of $k^2a^2(m_y^2+m_z^2)$. Unfortunately, the series has difficulty
with round-off errors as $k^2a^2(m_y^2+m_z^2)$ becomes large so we
only used the
integral over Bessel functions:
\begin{eqnarray}
q_m&\to &\pi\left(\frac{3\Gamma}{2\Omega ka} \right)^2\int_0^\infty
\frac{\rho |x|}{(x^2+\rho^2)^{3/2}}\left[ \frac{2x^2+\rho^2}{x^2+\rho^2}
J_0(s(\rho ))\right.\nonumber\\
&\null &\left. -\frac{\rho^2}{x^2+\rho^2}\frac{m_y^2-m_z^2}{m_y^2+m_z^2}
J_2(s(\rho ))\right] d\rho\\
s(\rho ) &=& k a \frac{\rho}{\sqrt{x^2+\rho^2}}\sqrt{m_y^2+m_z^2}
\end{eqnarray}
where the integration was done numerically. For large $|x|$, the
$q_m$ is independent of $x$. For circular polarization, the only
change is to drop the term with $J_2(s)$ because then the $q_m$
can only depend on $m_y^2+m_z^2$.

\section{Numerical methods}\label{sec:numapp}

While the various summations of parameters converge, their numerical
evaluation can be difficult if used in their simple form. In this
section, we give the numerical methods we used to evaluate various expressions.

The summation for ${\cal G}$ only slowly converges with system size. For
example, writing
\begin{eqnarray}\label{eq:gapp1}
{\cal G}(N)&=&\sum_{m\neq 0}^{m_y^2+m_z^2<N^2}g(\bm{R}_m)\\
\qquad {\cal G}&=& \lim_{N\to\infty}  {\cal G}(N)
\end{eqnarray}
only slowly converges because it is similar to integrating $\exp [ik\rho]/\rho$
in cylindrical coordinates. Thus, the hard limit to the summation gives
an effect similar to Gibbs phenomenon where the ${\cal G}(N)$ oscillates
around the limit.
Instead, we used a function to smoothly cut-off
the sum
\begin{equation}\label{eq:gapp2}
{\cal G}(N)=\sum_{m\neq 0}^{m_y^2+m_z^2<N^2}g(\bm{R}_m)e^{-36
(m_y^2+m_z^2)^2/N^4}
\end{equation}
with ${\cal G}$ being the limit as $N\to\infty$. As we increased $N$,
this expression smoothly approaches ${\cal G}$ so that the numerical
calculation is unambiguous. As a test, we compare the real part of
${\cal G}(N)$ to $(\Gamma /2)\{ (3/[4\pi ])(\lambda /a)^2-1\}$.
For example, for $a=0.8\lambda$, the fractional error from Eq.~(\ref{eq:gapp2})
of the real part of ${\cal G}$ was $1.1\times 10^{-9}$ for $N=125$,
$6.9\times 10^{-11}$ for 250, $4.3\times 10^{-12}$ for 500,
$2.7\times 10^{-13}$ for 1000, and $1.8\times 10^{-14}$ for 1500.
Whereas, Eq.~(\ref{eq:gapp1}) gives fractional errors
of 0.080 for $N=125$,
0.056 for 250, 0.040 for 500,
0.028 for 1000, and 0.023 for 1500.

For similar reasons, the summation for ${\cal \bar{G}}$ used in the
two array calculations does not converge well with system size. We
use the same kind of smooth cut-off 
\begin{equation}
{\cal \bar{G}}(N)=e^{-ikL}\sum_{m_\beta}^{m_y^2+m_z^2<N^2}
g(\bm{R}_{m_\beta})e^{-36
(m_y^2+m_z^2)^2/N^4}
\end{equation}
where $\bm{R}_{m_\beta}=(L,m_ya,m_za)$ and ${\cal \bar{G}}$ is the limit
as $N\to\infty$. Because of the $L$ in the position,
the $N$ has to be much larger than $L/a$ which sometimes meant using more
terms than needed
to converge ${\cal G}$. But as before, this expression smoothly
converges with $N$ giving an unambiguous value for smaller computational
cost. When $L>>\lambda$, the summation can be done analytically using
the same expression in Eq.~(\ref{eq:sumint1}) multiplied by $3\Gamma /4$
and $\exp (-ikL)$ which gives
\begin{equation}
{\cal \bar{G}}\to\frac{3\pi\Gamma}{2 k^2a^a} \qquad {\rm for} \qquad L\gg
\lambda
\end{equation}

In principle, the calculation of ${\cal Q}^{(2)}$, Eq.~(\ref{eq:q2mf2}),
does not need any
convergence help because both the $q_m$ and the cumulant decreases with
increasing $m_y^2+m_z^2$. In practice, the results converge faster with
a similar, smooth cut-off function. For MF2, the calculation extends
over terms with $\sqrt{m_y^2+m_z^2}\leq N_w$. In our calculations, we
did calculations with $N_w=15$, 20, 25, and 30. The calculation of
${\cal Q}^{(2)}$ converged faster when we used
\begin{equation}
{\cal Q}^{(2)}=\sum_{m\neq 0}^{m_y^2+m^2_z<N_w^2}q_m[\langle\hat{\sigma}^+_0\hat{\sigma}^-_m
\rangle -\langle\hat{\sigma}^+_0\rangle\langle\hat{\sigma}^-_0
\rangle ]W_m
\end{equation}
with
\begin{equation}
W_m = e^{-36 (m_y^2+m_z^2)^2/N_w^4}.
\end{equation}
As with ${\cal G}$, the weight function causes a smooth cutoff of the
cumulants times an oscillating term leading to faster convergence.

The calculation of the
${\cal C}^{\pm,0}_n$ parameters in the two atom equations
in Appendix~\ref{sec:twoat} converge in a similar way to the
${\cal G}$ parameters. However, we don't have the luxury of using
the same type of smooth cutoff
because the ${\cal C}^{\pm,0}_n$
contains expectation values of pair operators, neither at the
origin. For these parameters,
we used the sum as written.

\bibliography{two_inf_arr}

\begin{thebibliography}{48}%
\makeatletter
\providecommand \@ifxundefined [1]{%
 \@ifx{#1\undefined}
}%
\providecommand \@ifnum [1]{%
 \ifnum #1\expandafter \@firstoftwo
 \else \expandafter \@secondoftwo
 \fi
}%
\providecommand \@ifx [1]{%
 \ifx #1\expandafter \@firstoftwo
 \else \expandafter \@secondoftwo
 \fi
}%
\providecommand \natexlab [1]{#1}%
\providecommand \enquote  [1]{``#1''}%
\providecommand \bibnamefont  [1]{#1}%
\providecommand \bibfnamefont [1]{#1}%
\providecommand \citenamefont [1]{#1}%
\providecommand \href@noop [0]{\@secondoftwo}%
\providecommand \href [0]{\begingroup \@sanitize@url \@href}%
\providecommand \@href[1]{\@@startlink{#1}\@@href}%
\providecommand \@@href[1]{\endgroup#1\@@endlink}%
\providecommand \@sanitize@url [0]{\catcode `\\12\catcode `\$12\catcode
  `\&12\catcode `\#12\catcode `\^12\catcode `\_12\catcode `\%12\relax}%
\providecommand \@@startlink[1]{}%
\providecommand \@@endlink[0]{}%
\providecommand \url  [0]{\begingroup\@sanitize@url \@url }%
\providecommand \@url [1]{\endgroup\@href {#1}{\urlprefix }}%
\providecommand \urlprefix  [0]{URL }%
\providecommand \Eprint [0]{\href }%
\providecommand \doibase [0]{http://dx.doi.org/}%
\providecommand \selectlanguage [0]{\@gobble}%
\providecommand \bibinfo  [0]{\@secondoftwo}%
\providecommand \bibfield  [0]{\@secondoftwo}%
\providecommand \translation [1]{[#1]}%
\providecommand \BibitemOpen [0]{}%
\providecommand \bibitemStop [0]{}%
\providecommand \bibitemNoStop [0]{.\EOS\space}%
\providecommand \EOS [0]{\spacefactor3000\relax}%
\providecommand \BibitemShut  [1]{\csname bibitem#1\endcsname}%
\let\auto@bib@innerbib\@empty
\bibitem [{\citenamefont {Chang}\ \emph {et~al.}(2004)\citenamefont {Chang},
  \citenamefont {Ye},\ and\ \citenamefont {Lukin}}]{CYL2004}%
  \BibitemOpen
  \bibfield  {author} {\bibinfo {author} {\bibfnamefont {D.E.}\ \bibnamefont
  {Chang}}, \bibinfo {author} {\bibfnamefont {J.}~\bibnamefont {Ye}}, \ and\
  \bibinfo {author} {\bibfnamefont {M.D.}\ \bibnamefont {Lukin}},\ }\bibfield
  {title} {\enquote {\bibinfo {title} {Controlling dipole-dipole frequency
  shifts in a lattice-based optical atomic clock},}\ }\href@noop {} {\bibfield
  {journal} {\bibinfo  {journal} {Phys. Rev. A}\ }\textbf {\bibinfo {volume}
  {69}},\ \bibinfo {pages} {023810} (\bibinfo {year} {2004})}\BibitemShut
  {NoStop}%
\bibitem [{\citenamefont {Meir}\ \emph {et~al.}(2014)\citenamefont {Meir},
  \citenamefont {Schwartz}, \citenamefont {Shahmoon}, \citenamefont {Oron},\
  and\ \citenamefont {Ozeri}}]{MSS2014}%
  \BibitemOpen
  \bibfield  {author} {\bibinfo {author} {\bibfnamefont {Z.}~\bibnamefont
  {Meir}}, \bibinfo {author} {\bibfnamefont {O.}~\bibnamefont {Schwartz}},
  \bibinfo {author} {\bibfnamefont {E.}~\bibnamefont {Shahmoon}}, \bibinfo
  {author} {\bibfnamefont {D.}~\bibnamefont {Oron}}, \ and\ \bibinfo {author}
  {\bibfnamefont {R.}~\bibnamefont {Ozeri}},\ }\bibfield  {title} {\enquote
  {\bibinfo {title} {Cooperative {L}amb shift in a mesoscopic atomic array},}\
  }\href@noop {} {\bibfield  {journal} {\bibinfo  {journal} {Phys. Rev. Lett.}\
  }\textbf {\bibinfo {volume} {113}},\ \bibinfo {pages} {193002} (\bibinfo
  {year} {2014})}\BibitemShut {NoStop}%
\bibitem [{\citenamefont {Jenkins}\ and\ \citenamefont
  {Ruostekoski}(2012)}]{JR12012}%
  \BibitemOpen
  \bibfield  {author} {\bibinfo {author} {\bibfnamefont {S.~D.}\ \bibnamefont
  {Jenkins}}\ and\ \bibinfo {author} {\bibfnamefont {J.}~\bibnamefont
  {Ruostekoski}},\ }\bibfield  {title} {\enquote {\bibinfo {title} {Controlled
  manipulation of light by cooperative response of atoms in an optical
  lattice},}\ }\href@noop {} {\bibfield  {journal} {\bibinfo  {journal} {Phys.
  Rev. A}\ }\textbf {\bibinfo {volume} {86}},\ \bibinfo {pages} {031602(R)}
  (\bibinfo {year} {2012})}\BibitemShut {NoStop}%
\bibitem [{\citenamefont {Bettles}\ \emph {et~al.}(2016)\citenamefont
  {Bettles}, \citenamefont {Gardiner},\ and\ \citenamefont {Adams}}]{BGA2016}%
  \BibitemOpen
  \bibfield  {author} {\bibinfo {author} {\bibfnamefont {R.~J.}\ \bibnamefont
  {Bettles}}, \bibinfo {author} {\bibfnamefont {S.~A.}\ \bibnamefont
  {Gardiner}}, \ and\ \bibinfo {author} {\bibfnamefont {C.~S.}\ \bibnamefont
  {Adams}},\ }\bibfield  {title} {\enquote {\bibinfo {title} {Enhanced optical
  cross section via collective coupling of atomic dipoles in a 2d array},}\
  }\href@noop {} {\bibfield  {journal} {\bibinfo  {journal} {Phys. Rev. Lett.}\
  }\textbf {\bibinfo {volume} {116}},\ \bibinfo {pages} {103602} (\bibinfo
  {year} {2016})}\BibitemShut {NoStop}%
\bibitem [{\citenamefont {Sutherland}\ and\ \citenamefont
  {Robicheaux}(2016)}]{SR22016}%
  \BibitemOpen
  \bibfield  {author} {\bibinfo {author} {\bibfnamefont {R.~T.}\ \bibnamefont
  {Sutherland}}\ and\ \bibinfo {author} {\bibfnamefont {F.}~\bibnamefont
  {Robicheaux}},\ }\bibfield  {title} {\enquote {\bibinfo {title} {Collective
  dipole-dipole interactions in an atomic array},}\ }\href {\doibase
  10.1103/PhysRevA.94.013847} {\bibfield  {journal} {\bibinfo  {journal} {Phys.
  Rev. A}\ }\textbf {\bibinfo {volume} {94}},\ \bibinfo {pages} {013847}
  (\bibinfo {year} {2016})}\BibitemShut {NoStop}%
\bibitem [{\citenamefont {Facchinetti}\ \emph {et~al.}(2016)\citenamefont
  {Facchinetti}, \citenamefont {Jenkins},\ and\ \citenamefont
  {Ruostekoski}}]{FJR2016}%
  \BibitemOpen
  \bibfield  {author} {\bibinfo {author} {\bibfnamefont {G}~\bibnamefont
  {Facchinetti}}, \bibinfo {author} {\bibfnamefont {Stewart~D}\ \bibnamefont
  {Jenkins}}, \ and\ \bibinfo {author} {\bibfnamefont {Janne}\ \bibnamefont
  {Ruostekoski}},\ }\bibfield  {title} {\enquote {\bibinfo {title} {Storing
  light with subradiant correlations in arrays of atoms},}\ }\href@noop {}
  {\bibfield  {journal} {\bibinfo  {journal} {Phys. Rev. Lett.}\ }\textbf
  {\bibinfo {volume} {117}},\ \bibinfo {pages} {243601} (\bibinfo {year}
  {2016})}\BibitemShut {NoStop}%
\bibitem [{\citenamefont {Shahmoon}\ \emph {et~al.}(2017)\citenamefont
  {Shahmoon}, \citenamefont {Wild}, \citenamefont {Lukin},\ and\ \citenamefont
  {Yelin}}]{SWL2017}%
  \BibitemOpen
  \bibfield  {author} {\bibinfo {author} {\bibfnamefont {E.}~\bibnamefont
  {Shahmoon}}, \bibinfo {author} {\bibfnamefont {D.~S.}\ \bibnamefont {Wild}},
  \bibinfo {author} {\bibfnamefont {M.~D.}\ \bibnamefont {Lukin}}, \ and\
  \bibinfo {author} {\bibfnamefont {S.~F.}\ \bibnamefont {Yelin}},\ }\bibfield
  {title} {\enquote {\bibinfo {title} {Cooperative resonances in light
  scattering from two-dimensional atomic arrays},}\ }\href@noop {} {\bibfield
  {journal} {\bibinfo  {journal} {Phys. Rev. Lett.}\ }\textbf {\bibinfo
  {volume} {118}},\ \bibinfo {pages} {113601} (\bibinfo {year}
  {2017})}\BibitemShut {NoStop}%
\bibitem [{\citenamefont {Grankin}\ \emph {et~al.}(2018)\citenamefont
  {Grankin}, \citenamefont {Guimond}, \citenamefont {Vasilyev}, \citenamefont
  {Vermersch},\ and\ \citenamefont {Zoller}}]{GGV2018}%
  \BibitemOpen
  \bibfield  {author} {\bibinfo {author} {\bibfnamefont {A.}~\bibnamefont
  {Grankin}}, \bibinfo {author} {\bibfnamefont {P.-O.}\ \bibnamefont
  {Guimond}}, \bibinfo {author} {\bibfnamefont {D.~V.}\ \bibnamefont
  {Vasilyev}}, \bibinfo {author} {\bibfnamefont {B.}~\bibnamefont {Vermersch}},
  \ and\ \bibinfo {author} {\bibfnamefont {P.}~\bibnamefont {Zoller}},\
  }\bibfield  {title} {\enquote {\bibinfo {title} {Free-space photonic quantum
  link and chiral quantum optics},}\ }\href@noop {} {\bibfield  {journal}
  {\bibinfo  {journal} {Phys. Rev. A}\ }\textbf {\bibinfo {volume} {98}},\
  \bibinfo {pages} {043825} (\bibinfo {year} {2018})}\BibitemShut {NoStop}%
\bibitem [{\citenamefont {Chang}\ \emph {et~al.}(2018)\citenamefont {Chang},
  \citenamefont {Douglas}, \citenamefont {Gonz{\'a}lez-Tudela}, \citenamefont
  {Hung},\ and\ \citenamefont {Kimble}}]{CDG2018}%
  \BibitemOpen
  \bibfield  {author} {\bibinfo {author} {\bibfnamefont {D.E.}\ \bibnamefont
  {Chang}}, \bibinfo {author} {\bibfnamefont {J.S.}\ \bibnamefont {Douglas}},
  \bibinfo {author} {\bibfnamefont {A.}~\bibnamefont {Gonz{\'a}lez-Tudela}},
  \bibinfo {author} {\bibfnamefont {C.-.L}\ \bibnamefont {Hung}}, \ and\
  \bibinfo {author} {\bibfnamefont {H.J.}\ \bibnamefont {Kimble}},\ }\bibfield
  {title} {\enquote {\bibinfo {title} {Colloquium: Quantum matter built from
  nanoscopic lattices of atoms and photons},}\ }\href@noop {} {\bibfield
  {journal} {\bibinfo  {journal} {Rev. Mod. Phys.}\ }\textbf {\bibinfo {volume}
  {90}},\ \bibinfo {pages} {031002} (\bibinfo {year} {2018})}\BibitemShut
  {NoStop}%
\bibitem [{\citenamefont {Asenjo-Garcia}\ \emph {et~al.}(2017)\citenamefont
  {Asenjo-Garcia}, \citenamefont {Moreno-Cardoner}, \citenamefont {Albrecht},
  \citenamefont {Kimble},\ and\ \citenamefont {Chang}}]{AMA2017}%
  \BibitemOpen
  \bibfield  {author} {\bibinfo {author} {\bibfnamefont {A.}~\bibnamefont
  {Asenjo-Garcia}}, \bibinfo {author} {\bibfnamefont {M.}~\bibnamefont
  {Moreno-Cardoner}}, \bibinfo {author} {\bibfnamefont {A.}~\bibnamefont
  {Albrecht}}, \bibinfo {author} {\bibfnamefont {H.~J.}\ \bibnamefont
  {Kimble}}, \ and\ \bibinfo {author} {\bibfnamefont {D.~E.}\ \bibnamefont
  {Chang}},\ }\bibfield  {title} {\enquote {\bibinfo {title} {Exponential
  improvement in photon storage fidelities using subradiance and “selective
  radiance” in atomic arrays},}\ }\href@noop {} {\bibfield  {journal}
  {\bibinfo  {journal} {Phys. Rev. X}\ }\textbf {\bibinfo {volume} {7}},\
  \bibinfo {pages} {031024} (\bibinfo {year} {2017})}\BibitemShut {NoStop}%
\bibitem [{\citenamefont {Asenjo-Garcia}\ \emph {et~al.}(2019)\citenamefont
  {Asenjo-Garcia}, \citenamefont {Kimble},\ and\ \citenamefont
  {Chang}}]{AKC2019}%
  \BibitemOpen
  \bibfield  {author} {\bibinfo {author} {\bibfnamefont {A.}~\bibnamefont
  {Asenjo-Garcia}}, \bibinfo {author} {\bibfnamefont {H.~J.}\ \bibnamefont
  {Kimble}}, \ and\ \bibinfo {author} {\bibfnamefont {D.~E.}\ \bibnamefont
  {Chang}},\ }\bibfield  {title} {\enquote {\bibinfo {title} {Optical
  waveguiding by atomic entanglement in multilevel atom arrays},}\ }\href@noop
  {} {\bibfield  {journal} {\bibinfo  {journal} {Proc. Nat. Ac. Sci.}\ }\textbf
  {\bibinfo {volume} {116}},\ \bibinfo {pages} {25503} (\bibinfo {year}
  {2019})}\BibitemShut {NoStop}%
\bibitem [{\citenamefont {Henriet}\ \emph {et~al.}(2019)\citenamefont
  {Henriet}, \citenamefont {Douglas}, \citenamefont {Chang},\ and\
  \citenamefont {Albrecht}}]{HDC2019}%
  \BibitemOpen
  \bibfield  {author} {\bibinfo {author} {\bibfnamefont {L.}~\bibnamefont
  {Henriet}}, \bibinfo {author} {\bibfnamefont {J.~S.}\ \bibnamefont
  {Douglas}}, \bibinfo {author} {\bibfnamefont {D.~E.}\ \bibnamefont {Chang}},
  \ and\ \bibinfo {author} {\bibfnamefont {A.}~\bibnamefont {Albrecht}},\
  }\bibfield  {title} {\enquote {\bibinfo {title} {Critical open-system
  dynamics in a one-dimensional optical-lattice clock},}\ }\href@noop {}
  {\bibfield  {journal} {\bibinfo  {journal} {Phy. Rev. A}\ }\textbf {\bibinfo
  {volume} {99}},\ \bibinfo {pages} {023802} (\bibinfo {year}
  {2019})}\BibitemShut {NoStop}%
\bibitem [{\citenamefont {Needham}\ \emph {et~al.}(2019)\citenamefont
  {Needham}, \citenamefont {Lesanovsky},\ and\ \citenamefont
  {Olmos}}]{NLO2019}%
  \BibitemOpen
  \bibfield  {author} {\bibinfo {author} {\bibfnamefont {J.~A.}\ \bibnamefont
  {Needham}}, \bibinfo {author} {\bibfnamefont {I.}~\bibnamefont {Lesanovsky}},
  \ and\ \bibinfo {author} {\bibfnamefont {B.}~\bibnamefont {Olmos}},\
  }\bibfield  {title} {\enquote {\bibinfo {title} {Subradiance-protected
  excitation transport},}\ }\href@noop {} {\bibfield  {journal} {\bibinfo
  {journal} {New J. Phys.}\ }\textbf {\bibinfo {volume} {21}},\ \bibinfo
  {pages} {073061} (\bibinfo {year} {2019})}\BibitemShut {NoStop}%
\bibitem [{\citenamefont {Qu}\ and\ \citenamefont {Rey}(2019)}]{QR12019}%
  \BibitemOpen
  \bibfield  {author} {\bibinfo {author} {\bibfnamefont {C.}~\bibnamefont
  {Qu}}\ and\ \bibinfo {author} {\bibfnamefont {A.~M.}\ \bibnamefont {Rey}},\
  }\bibfield  {title} {\enquote {\bibinfo {title} {Spin squeezing and many-body
  dipolar dynamics in optical lattice clocks},}\ }\href@noop {} {\bibfield
  {journal} {\bibinfo  {journal} {Phys. Rev. A}\ }\textbf {\bibinfo {volume}
  {100}},\ \bibinfo {pages} {041602} (\bibinfo {year} {2019})}\BibitemShut
  {NoStop}%
\bibitem [{\citenamefont {Zhang}\ and\ \citenamefont
  {M{\o}lmer}(2019)}]{ZM12019}%
  \BibitemOpen
  \bibfield  {author} {\bibinfo {author} {\bibfnamefont {Yu-Xiang}\
  \bibnamefont {Zhang}}\ and\ \bibinfo {author} {\bibfnamefont {Klaus}\
  \bibnamefont {M{\o}lmer}},\ }\bibfield  {title} {\enquote {\bibinfo {title}
  {Theory of subradiant states of a one-dimensional two-level atom chain},}\
  }\href@noop {} {\bibfield  {journal} {\bibinfo  {journal} {Phys. Rev. lett.}\
  }\textbf {\bibinfo {volume} {122}},\ \bibinfo {pages} {203605} (\bibinfo
  {year} {2019})}\BibitemShut {NoStop}%
\bibitem [{\citenamefont {Masson}\ \emph {et~al.}(2020)\citenamefont {Masson},
  \citenamefont {Ferrier-Barbut}, \citenamefont {Orozco}, \citenamefont
  {Browaeys},\ and\ \citenamefont {Asenjo-Garcia}}]{MFO2020}%
  \BibitemOpen
  \bibfield  {author} {\bibinfo {author} {\bibfnamefont {S.~J.}\ \bibnamefont
  {Masson}}, \bibinfo {author} {\bibfnamefont {I.}~\bibnamefont
  {Ferrier-Barbut}}, \bibinfo {author} {\bibfnamefont {L.~A.}\ \bibnamefont
  {Orozco}}, \bibinfo {author} {\bibfnamefont {A.}~\bibnamefont {Browaeys}}, \
  and\ \bibinfo {author} {\bibfnamefont {A.}~\bibnamefont {Asenjo-Garcia}},\
  }\bibfield  {title} {\enquote {\bibinfo {title} {Many-body signatures of
  collective decay in atomic chains},}\ }\href@noop {} {\bibfield  {journal}
  {\bibinfo  {journal} {Phys. Rev. Lett.}\ }\textbf {\bibinfo {volume} {125}},\
  \bibinfo {pages} {263601} (\bibinfo {year} {2020})}\BibitemShut {NoStop}%
\bibitem [{\citenamefont {Javanainen}\ and\ \citenamefont
  {Rajapakse}(2019)}]{JR12019}%
  \BibitemOpen
  \bibfield  {author} {\bibinfo {author} {\bibfnamefont {J.}~\bibnamefont
  {Javanainen}}\ and\ \bibinfo {author} {\bibfnamefont {R.}~\bibnamefont
  {Rajapakse}},\ }\bibfield  {title} {\enquote {\bibinfo {title} {Light
  propagation in systems involving two-dimensional atomic lattices},}\
  }\href@noop {} {\bibfield  {journal} {\bibinfo  {journal} {Phys. Rev. A}\
  }\textbf {\bibinfo {volume} {100}},\ \bibinfo {pages} {013616} (\bibinfo
  {year} {2019})}\BibitemShut {NoStop}%
\bibitem [{\citenamefont {Williamson}\ \emph {et~al.}(2020)\citenamefont
  {Williamson}, \citenamefont {Borgh},\ and\ \citenamefont
  {Ruostekoski}}]{WBR2020}%
  \BibitemOpen
  \bibfield  {author} {\bibinfo {author} {\bibfnamefont {L.~A.}\ \bibnamefont
  {Williamson}}, \bibinfo {author} {\bibfnamefont {M.~O.}\ \bibnamefont
  {Borgh}}, \ and\ \bibinfo {author} {\bibfnamefont {J.}~\bibnamefont
  {Ruostekoski}},\ }\bibfield  {title} {\enquote {\bibinfo {title} {Superatom
  picture of collective nonclassical light emission and dipole blockade in atom
  arrays},}\ }\href@noop {} {\bibfield  {journal} {\bibinfo  {journal} {Phys.
  Rev. Lett.}\ }\textbf {\bibinfo {volume} {125}},\ \bibinfo {pages} {073602}
  (\bibinfo {year} {2020})}\BibitemShut {NoStop}%
\bibitem [{\citenamefont {Williamson}\ and\ \citenamefont
  {Ruostekoski}(2020)}]{WR12020}%
  \BibitemOpen
  \bibfield  {author} {\bibinfo {author} {\bibfnamefont {L.~A.}\ \bibnamefont
  {Williamson}}\ and\ \bibinfo {author} {\bibfnamefont {J.}~\bibnamefont
  {Ruostekoski}},\ }\bibfield  {title} {\enquote {\bibinfo {title} {Optical
  response of atom chains beyond the limit of low light intensity: {T}he
  validity of the linear classical oscillator model},}\ }\href@noop {}
  {\bibfield  {journal} {\bibinfo  {journal} {Phys. Rev. Research}\ }\textbf
  {\bibinfo {volume} {2}},\ \bibinfo {pages} {023273} (\bibinfo {year}
  {2020})}\BibitemShut {NoStop}%
\bibitem [{\citenamefont {Bettles}\ \emph {et~al.}(2020)\citenamefont
  {Bettles}, \citenamefont {Lee}, \citenamefont {Gardiner},\ and\ \citenamefont
  {Ruostekoski}}]{BLG2020}%
  \BibitemOpen
  \bibfield  {author} {\bibinfo {author} {\bibfnamefont {Robert~J}\
  \bibnamefont {Bettles}}, \bibinfo {author} {\bibfnamefont {Mark~D}\
  \bibnamefont {Lee}}, \bibinfo {author} {\bibfnamefont {Simon~A}\ \bibnamefont
  {Gardiner}}, \ and\ \bibinfo {author} {\bibfnamefont {Janne}\ \bibnamefont
  {Ruostekoski}},\ }\bibfield  {title} {\enquote {\bibinfo {title} {Quantum and
  nonlinear effects in light transmitted through planar atomic arrays},}\
  }\href@noop {} {\bibfield  {journal} {\bibinfo  {journal} {Commun. Phys.}\
  }\textbf {\bibinfo {volume} {3}},\ \bibinfo {pages} {141} (\bibinfo {year}
  {2020})}\BibitemShut {NoStop}%
\bibitem [{\citenamefont {Guimond}\ \emph {et~al.}(2019)\citenamefont
  {Guimond}, \citenamefont {Grankin}, \citenamefont {Vasilyev}, \citenamefont
  {Vermersch},\ and\ \citenamefont {Zoller}}]{GGV2019}%
  \BibitemOpen
  \bibfield  {author} {\bibinfo {author} {\bibfnamefont {P.-O.}\ \bibnamefont
  {Guimond}}, \bibinfo {author} {\bibfnamefont {A.}~\bibnamefont {Grankin}},
  \bibinfo {author} {\bibfnamefont {D.~V.}\ \bibnamefont {Vasilyev}}, \bibinfo
  {author} {\bibfnamefont {B.}~\bibnamefont {Vermersch}}, \ and\ \bibinfo
  {author} {\bibfnamefont {P.}~\bibnamefont {Zoller}},\ }\bibfield  {title}
  {\enquote {\bibinfo {title} {Subradiant {B}ell states in distant atomic
  arrays},}\ }\href@noop {} {\bibfield  {journal} {\bibinfo  {journal} {Phys.
  Rev. Lett.}\ }\textbf {\bibinfo {volume} {122}},\ \bibinfo {pages} {093601}
  (\bibinfo {year} {2019})}\BibitemShut {NoStop}%
\bibitem [{\citenamefont {Ballantine}\ and\ \citenamefont
  {Ruostekoski}(2020)}]{BR12020}%
  \BibitemOpen
  \bibfield  {author} {\bibinfo {author} {\bibfnamefont {K.E.}\ \bibnamefont
  {Ballantine}}\ and\ \bibinfo {author} {\bibfnamefont {J.}~\bibnamefont
  {Ruostekoski}},\ }\bibfield  {title} {\enquote {\bibinfo {title} {Optical
  magnetism and {H}uygens’ surfaces in arrays of atoms induced by cooperative
  responses},}\ }\href@noop {} {\bibfield  {journal} {\bibinfo  {journal}
  {Phys. Rev. Lett.}\ }\textbf {\bibinfo {volume} {125}},\ \bibinfo {pages}
  {143604} (\bibinfo {year} {2020})}\BibitemShut {NoStop}%
\bibitem [{\citenamefont {Bekenstein}\ \emph {et~al.}(2020)\citenamefont
  {Bekenstein}, \citenamefont {Pikovski}, \citenamefont {Pichler},
  \citenamefont {Shahmoon}, \citenamefont {Yelin},\ and\ \citenamefont
  {Lukin}}]{BPP2020}%
  \BibitemOpen
  \bibfield  {author} {\bibinfo {author} {\bibfnamefont {R.}~\bibnamefont
  {Bekenstein}}, \bibinfo {author} {\bibfnamefont {I.}~\bibnamefont
  {Pikovski}}, \bibinfo {author} {\bibfnamefont {H.}~\bibnamefont {Pichler}},
  \bibinfo {author} {\bibfnamefont {E.}~\bibnamefont {Shahmoon}}, \bibinfo
  {author} {\bibfnamefont {S.~F.}\ \bibnamefont {Yelin}}, \ and\ \bibinfo
  {author} {\bibfnamefont {M.~D.}\ \bibnamefont {Lukin}},\ }\bibfield  {title}
  {\enquote {\bibinfo {title} {Quantum metasurfaces with atom arrays},}\
  }\href@noop {} {\bibfield  {journal} {\bibinfo  {journal} {Nature Phys.}\
  }\textbf {\bibinfo {volume} {16}},\ \bibinfo {pages} {676} (\bibinfo {year}
  {2020})}\BibitemShut {NoStop}%
\bibitem [{\citenamefont {Rui}\ \emph {et~al.}(2020)\citenamefont {Rui},
  \citenamefont {Wei}, \citenamefont {Rubio-Abadal}, \citenamefont {Hollerith},
  \citenamefont {Zeiher}, \citenamefont {Stamper-Kurn}, \citenamefont {Gross},\
  and\ \citenamefont {Bloch}}]{RWR2020}%
  \BibitemOpen
  \bibfield  {author} {\bibinfo {author} {\bibfnamefont {Jun}\ \bibnamefont
  {Rui}}, \bibinfo {author} {\bibfnamefont {D.}~\bibnamefont {Wei}}, \bibinfo
  {author} {\bibfnamefont {A.}~\bibnamefont {Rubio-Abadal}}, \bibinfo {author}
  {\bibfnamefont {S.}~\bibnamefont {Hollerith}}, \bibinfo {author}
  {\bibfnamefont {J.}~\bibnamefont {Zeiher}}, \bibinfo {author} {\bibfnamefont
  {D.~M.}\ \bibnamefont {Stamper-Kurn}}, \bibinfo {author} {\bibfnamefont
  {C.}~\bibnamefont {Gross}}, \ and\ \bibinfo {author} {\bibfnamefont
  {I.}~\bibnamefont {Bloch}},\ }\bibfield  {title} {\enquote {\bibinfo {title}
  {A subradiant optical mirror formed by a single structured atomic layer},}\
  }\href@noop {} {\bibfield  {journal} {\bibinfo  {journal} {Nature}\ }\textbf
  {\bibinfo {volume} {583}},\ \bibinfo {pages} {369} (\bibinfo {year}
  {2020})}\BibitemShut {NoStop}%
\bibitem [{\citenamefont {Cidrim}\ \emph {et~al.}(2020)\citenamefont {Cidrim},
  \citenamefont {do~Espirito~Santo}, \citenamefont {Schachenmayer},
  \citenamefont {Kaiser},\ and\ \citenamefont {Bachelard}}]{CES2020}%
  \BibitemOpen
  \bibfield  {author} {\bibinfo {author} {\bibfnamefont {A.}~\bibnamefont
  {Cidrim}}, \bibinfo {author} {\bibfnamefont {T.~S.}\ \bibnamefont
  {do~Espirito~Santo}}, \bibinfo {author} {\bibfnamefont {J.}~\bibnamefont
  {Schachenmayer}}, \bibinfo {author} {\bibfnamefont {R.}~\bibnamefont
  {Kaiser}}, \ and\ \bibinfo {author} {\bibfnamefont {R.}~\bibnamefont
  {Bachelard}},\ }\bibfield  {title} {\enquote {\bibinfo {title} {Photon
  blockade with ground-state neutral atoms},}\ }\href@noop {} {\bibfield
  {journal} {\bibinfo  {journal} {Phys. Rev. Lett.}\ }\textbf {\bibinfo
  {volume} {125}},\ \bibinfo {pages} {073601} (\bibinfo {year}
  {2020})}\BibitemShut {NoStop}%
\bibitem [{\citenamefont {Ballantine}\ and\ \citenamefont
  {Ruostekoski}(2021)}]{BR12021}%
  \BibitemOpen
  \bibfield  {author} {\bibinfo {author} {\bibfnamefont {K.~E.}\ \bibnamefont
  {Ballantine}}\ and\ \bibinfo {author} {\bibfnamefont {J.}~\bibnamefont
  {Ruostekoski}},\ }\bibfield  {title} {\enquote {\bibinfo {title} {Cooperative
  optical wavefront engineering with atomic arrays},}\ }\href@noop {}
  {\bibfield  {journal} {\bibinfo  {journal} {Nanophotonics}\ }\textbf
  {\bibinfo {volume} {10}},\ \bibinfo {pages} {1901} (\bibinfo {year}
  {2021})}\BibitemShut {NoStop}%
\bibitem [{\citenamefont {Cano}(2021)}]{DC12021}%
  \BibitemOpen
  \bibfield  {author} {\bibinfo {author} {\bibfnamefont {D.}~\bibnamefont
  {Cano}},\ }\bibfield  {title} {\enquote {\bibinfo {title} {Photon statistics
  of the light transmitted and reflected by a two-dimensional atomic array},}\
  }\href@noop {} {\bibfield  {journal} {\bibinfo  {journal} {Phys. Rev. A}\
  }\textbf {\bibinfo {volume} {104}},\ \bibinfo {pages} {053709} (\bibinfo
  {year} {2021})}\BibitemShut {NoStop}%
\bibitem [{\citenamefont {Ballantine}\ and\ \citenamefont
  {Ruostekoski}(2022)}]{BR12022}%
  \BibitemOpen
  \bibfield  {author} {\bibinfo {author} {\bibfnamefont {K.~E.}\ \bibnamefont
  {Ballantine}}\ and\ \bibinfo {author} {\bibfnamefont {J.}~\bibnamefont
  {Ruostekoski}},\ }\bibfield  {title} {\enquote {\bibinfo {title}
  {Unidirectional absorption, storage, and emission of single photons in a
  collectively responding bilayer atomic array},}\ }\href@noop {} {\bibfield
  {journal} {\bibinfo  {journal} {Phys. Rev. Research}\ }\textbf {\bibinfo
  {volume} {4}},\ \bibinfo {pages} {033200} (\bibinfo {year}
  {2022})}\BibitemShut {NoStop}%
\bibitem [{\citenamefont {Pedersen}\ \emph {et~al.}(2023)\citenamefont
  {Pedersen}, \citenamefont {Zhang},\ and\ \citenamefont {Pohl}}]{PZP2023}%
  \BibitemOpen
  \bibfield  {author} {\bibinfo {author} {\bibfnamefont {S.~P.}\ \bibnamefont
  {Pedersen}}, \bibinfo {author} {\bibfnamefont {L.}~\bibnamefont {Zhang}}, \
  and\ \bibinfo {author} {\bibfnamefont {T.}~\bibnamefont {Pohl}},\ }\bibfield
  {title} {\enquote {\bibinfo {title} {Quantum nonlinear metasurfaces from dual
  arrays of ultracold atoms},}\ }\href@noop {} {\bibfield  {journal} {\bibinfo
  {journal} {Phys. Rev. Research}\ }\textbf {\bibinfo {volume} {5}},\ \bibinfo
  {pages} {L012047} (\bibinfo {year} {2023})}\BibitemShut {NoStop}%
\bibitem [{\citenamefont {Patti}\ \emph {et~al.}(2021)\citenamefont {Patti},
  \citenamefont {Wild}, \citenamefont {Shahmoon}, \citenamefont {Lukin},\ and\
  \citenamefont {Yelin}}]{PWS2021}%
  \BibitemOpen
  \bibfield  {author} {\bibinfo {author} {\bibfnamefont {T.~L.}\ \bibnamefont
  {Patti}}, \bibinfo {author} {\bibfnamefont {D.~S.}\ \bibnamefont {Wild}},
  \bibinfo {author} {\bibfnamefont {E.}~\bibnamefont {Shahmoon}}, \bibinfo
  {author} {\bibfnamefont {M.~D.}\ \bibnamefont {Lukin}}, \ and\ \bibinfo
  {author} {\bibfnamefont {S.~F.}\ \bibnamefont {Yelin}},\ }\bibfield  {title}
  {\enquote {\bibinfo {title} {Controlling interactions between quantum
  emitters using atom arrays},}\ }\href@noop {} {\bibfield  {journal} {\bibinfo
   {journal} {Phys. Rev. Lett.}\ }\textbf {\bibinfo {volume} {126}},\ \bibinfo
  {pages} {223602} (\bibinfo {year} {2021})}\BibitemShut {NoStop}%
\bibitem [{\citenamefont {Robicheaux}(2021)}]{FR12021}%
  \BibitemOpen
  \bibfield  {author} {\bibinfo {author} {\bibfnamefont {F.}~\bibnamefont
  {Robicheaux}},\ }\bibfield  {title} {\enquote {\bibinfo {title} {Theoretical
  study of early-time superradiance for atom clouds and arrays},}\ }\href@noop
  {} {\bibfield  {journal} {\bibinfo  {journal} {Phys. Rev. A}\ }\textbf
  {\bibinfo {volume} {104}},\ \bibinfo {pages} {063706} (\bibinfo {year}
  {2021})}\BibitemShut {NoStop}%
\bibitem [{\citenamefont {Rubies-Bigorda}\ and\ \citenamefont
  {Yelin}(2022)}]{RBY2022}%
  \BibitemOpen
  \bibfield  {author} {\bibinfo {author} {\bibfnamefont {O.}~\bibnamefont
  {Rubies-Bigorda}}\ and\ \bibinfo {author} {\bibfnamefont {S.~.F}\
  \bibnamefont {Yelin}},\ }\bibfield  {title} {\enquote {\bibinfo {title}
  {Superradiance and subradiance in inverted atomic arrays},}\ }\href@noop {}
  {\bibfield  {journal} {\bibinfo  {journal} {Phys. Rev. A}\ }\textbf {\bibinfo
  {volume} {106}},\ \bibinfo {pages} {053717} (\bibinfo {year}
  {2022})}\BibitemShut {NoStop}%
\bibitem [{\citenamefont {Masson}\ and\ \citenamefont
  {Asenjo-Garcia}(2022)}]{MAG2022}%
  \BibitemOpen
  \bibfield  {author} {\bibinfo {author} {\bibfnamefont {S.~J.}\ \bibnamefont
  {Masson}}\ and\ \bibinfo {author} {\bibfnamefont {A.}~\bibnamefont
  {Asenjo-Garcia}},\ }\bibfield  {title} {\enquote {\bibinfo {title}
  {Universality of {D}icke superradiance in arrays of quantum emitters},}\
  }\href@noop {} {\bibfield  {journal} {\bibinfo  {journal} {Nature Comm.}\
  }\textbf {\bibinfo {volume} {13}},\ \bibinfo {pages} {2285} (\bibinfo {year}
  {2022})}\BibitemShut {NoStop}%
\bibitem [{\citenamefont {Sierra}\ \emph {et~al.}(2022)\citenamefont {Sierra},
  \citenamefont {Masson},\ and\ \citenamefont {Asenjo-Garcia}}]{SMA2022}%
  \BibitemOpen
  \bibfield  {author} {\bibinfo {author} {\bibfnamefont {E.}~\bibnamefont
  {Sierra}}, \bibinfo {author} {\bibfnamefont {S.~J.}\ \bibnamefont {Masson}},
  \ and\ \bibinfo {author} {\bibfnamefont {A.}~\bibnamefont {Asenjo-Garcia}},\
  }\bibfield  {title} {\enquote {\bibinfo {title} {Dicke superradiance in
  ordered lattices: dimensionality matters},}\ }\href@noop {} {\bibfield
  {journal} {\bibinfo  {journal} {Phys. Rev. Research}\ }\textbf {\bibinfo
  {volume} {4}},\ \bibinfo {pages} {023207} (\bibinfo {year}
  {2022})}\BibitemShut {NoStop}%
\bibitem [{\citenamefont {Rubies-Bigorda}\ \emph {et~al.}(2023)\citenamefont
  {Rubies-Bigorda}, \citenamefont {Ostermann},\ and\ \citenamefont
  {Yelin}}]{ROY2023}%
  \BibitemOpen
  \bibfield  {author} {\bibinfo {author} {\bibfnamefont {O.}~\bibnamefont
  {Rubies-Bigorda}}, \bibinfo {author} {\bibfnamefont {S.}~\bibnamefont
  {Ostermann}}, \ and\ \bibinfo {author} {\bibfnamefont {S.~F.}\ \bibnamefont
  {Yelin}},\ }\bibfield  {title} {\enquote {\bibinfo {title} {Characterizing
  superradiant dynamics in atomic arrays via a cumulant expansion approach},}\
  }\href@noop {} {\bibfield  {journal} {\bibinfo  {journal} {Phys. Rev.
  Research}\ }\textbf {\bibinfo {volume} {5}},\ \bibinfo {pages} {013091}
  (\bibinfo {year} {2023})}\BibitemShut {NoStop}%
\bibitem [{\citenamefont {Rubies-Bigorda}\ \emph {et~al.}(2022)\citenamefont
  {Rubies-Bigorda}, \citenamefont {Ostermann},\ and\ \citenamefont
  {Yelin}}]{ROYb2023}%
  \BibitemOpen
  \bibfield  {author} {\bibinfo {author} {\bibfnamefont {O.}~\bibnamefont
  {Rubies-Bigorda}}, \bibinfo {author} {\bibfnamefont {S.}~\bibnamefont
  {Ostermann}}, \ and\ \bibinfo {author} {\bibfnamefont {S.~F.}\ \bibnamefont
  {Yelin}},\ }\bibfield  {title} {\enquote {\bibinfo {title} {Generating
  multi-excitation subradiant states in incoherently excited atomic arrays},}\
  }\href@noop {} {\bibfield  {journal} {\bibinfo  {journal} {arXiv preprint
  arXiv:2209.00034}\ } (\bibinfo {year} {2022})}\BibitemShut {NoStop}%
\bibitem [{\citenamefont {Robicheaux}\ and\ \citenamefont
  {Suresh}(2021)}]{RS12021}%
  \BibitemOpen
  \bibfield  {author} {\bibinfo {author} {\bibfnamefont {F.}~\bibnamefont
  {Robicheaux}}\ and\ \bibinfo {author} {\bibfnamefont {D.~A.}\ \bibnamefont
  {Suresh}},\ }\bibfield  {title} {\enquote {\bibinfo {title} {Beyond lowest
  order mean-field theory for light interacting with atom arrays},}\
  }\href@noop {} {\bibfield  {journal} {\bibinfo  {journal} {Phys. Rev. A}\
  }\textbf {\bibinfo {volume} {104}},\ \bibinfo {pages} {023702} (\bibinfo
  {year} {2021})}\BibitemShut {NoStop}%
\bibitem [{\citenamefont {Kubo}(1962)}]{RK11962}%
  \BibitemOpen
  \bibfield  {author} {\bibinfo {author} {\bibfnamefont {R.}~\bibnamefont
  {Kubo}},\ }\bibfield  {title} {\enquote {\bibinfo {title} {Generalized
  cumulant expansion method},}\ }\href@noop {} {\bibfield  {journal} {\bibinfo
  {journal} {Journal of the Physical Society of Japan}\ }\textbf {\bibinfo
  {volume} {17}},\ \bibinfo {pages} {1100--1120} (\bibinfo {year}
  {1962})}\BibitemShut {NoStop}%
\bibitem [{\citenamefont {Fleischhauer}\ and\ \citenamefont
  {Yelin}(1999)}]{FY11999}%
  \BibitemOpen
  \bibfield  {author} {\bibinfo {author} {\bibfnamefont {Michael}\ \bibnamefont
  {Fleischhauer}}\ and\ \bibinfo {author} {\bibfnamefont {Susanne~F}\
  \bibnamefont {Yelin}},\ }\bibfield  {title} {\enquote {\bibinfo {title}
  {Radiative atom-atom interactions in optically dense media: Quantum
  corrections to the {L}orentz-{L}orenz formula},}\ }\href@noop {} {\bibfield
  {journal} {\bibinfo  {journal} {Phys. Rev. A}\ }\textbf {\bibinfo {volume}
  {59}},\ \bibinfo {pages} {2427} (\bibinfo {year} {1999})}\BibitemShut
  {NoStop}%
\bibitem [{\citenamefont {Lin}\ and\ \citenamefont {Yelin}(2012)}]{LY12012}%
  \BibitemOpen
  \bibfield  {author} {\bibinfo {author} {\bibfnamefont {G.-D.}\ \bibnamefont
  {Lin}}\ and\ \bibinfo {author} {\bibfnamefont {S.~F.}\ \bibnamefont
  {Yelin}},\ }\bibfield  {title} {\enquote {\bibinfo {title} {Superradiance in
  spin-$j$ particles: Effects of multiple levels},}\ }\href {\doibase
  10.1103/PhysRevA.85.033831} {\bibfield  {journal} {\bibinfo  {journal} {Phys.
  Rev. A}\ }\textbf {\bibinfo {volume} {85}},\ \bibinfo {pages} {033831}
  (\bibinfo {year} {2012})}\BibitemShut {NoStop}%
\bibitem [{\citenamefont {Kr{\"a}mer}\ and\ \citenamefont
  {Ritsch}(2015)}]{KR12015}%
  \BibitemOpen
  \bibfield  {author} {\bibinfo {author} {\bibfnamefont {S.}~\bibnamefont
  {Kr{\"a}mer}}\ and\ \bibinfo {author} {\bibfnamefont {H.}~\bibnamefont
  {Ritsch}},\ }\bibfield  {title} {\enquote {\bibinfo {title} {Generalized
  mean-field approach to simulate the dynamics of large open spin ensembles
  with long range interactions},}\ }\href@noop {} {\bibfield  {journal}
  {\bibinfo  {journal} {Eur. Phys. J. D}\ }\textbf {\bibinfo {volume} {69}},\
  \bibinfo {pages} {282} (\bibinfo {year} {2015})}\BibitemShut {NoStop}%
\bibitem [{\citenamefont {Kirton}\ and\ \citenamefont
  {Keeling}(2018)}]{KK12018}%
  \BibitemOpen
  \bibfield  {author} {\bibinfo {author} {\bibfnamefont {P.}~\bibnamefont
  {Kirton}}\ and\ \bibinfo {author} {\bibfnamefont {J.}~\bibnamefont
  {Keeling}},\ }\bibfield  {title} {\enquote {\bibinfo {title} {Superradiant
  and lasing states in driven-dissipative {D}icke models},}\ }\href@noop {}
  {\bibfield  {journal} {\bibinfo  {journal} {New J. Phys.}\ }\textbf {\bibinfo
  {volume} {20}},\ \bibinfo {pages} {015009} (\bibinfo {year}
  {2018})}\BibitemShut {NoStop}%
\bibitem [{\citenamefont {Kirton}\ \emph {et~al.}(2019)\citenamefont {Kirton},
  \citenamefont {Roses}, \citenamefont {Keeling},\ and\ \citenamefont
  {Dalla~Torre}}]{KRK2019}%
  \BibitemOpen
  \bibfield  {author} {\bibinfo {author} {\bibfnamefont {P.}~\bibnamefont
  {Kirton}}, \bibinfo {author} {\bibfnamefont {M.~M.}\ \bibnamefont {Roses}},
  \bibinfo {author} {\bibfnamefont {J.}~\bibnamefont {Keeling}}, \ and\
  \bibinfo {author} {\bibfnamefont {E.~G.}\ \bibnamefont {Dalla~Torre}},\
  }\bibfield  {title} {\enquote {\bibinfo {title} {Introduction to the {D}icke
  model: from equilibrium to nonequilibrium, and vice versa},}\ }\href@noop {}
  {\bibfield  {journal} {\bibinfo  {journal} {Adv. Quan. Tech.}\ }\textbf
  {\bibinfo {volume} {2}},\ \bibinfo {pages} {1800043} (\bibinfo {year}
  {2019})}\BibitemShut {NoStop}%
\bibitem [{\citenamefont {Ostermann}\ \emph {et~al.}(2019)\citenamefont
  {Ostermann}, \citenamefont {Meignant}, \citenamefont {Genes},\ and\
  \citenamefont {Ritsch}}]{OMG2019}%
  \BibitemOpen
  \bibfield  {author} {\bibinfo {author} {\bibfnamefont {L.}~\bibnamefont
  {Ostermann}}, \bibinfo {author} {\bibfnamefont {C.}~\bibnamefont {Meignant}},
  \bibinfo {author} {\bibfnamefont {C.}~\bibnamefont {Genes}}, \ and\ \bibinfo
  {author} {\bibfnamefont {H.}~\bibnamefont {Ritsch}},\ }\bibfield  {title}
  {\enquote {\bibinfo {title} {Super-and subradiance of clock atoms in
  multimode optical waveguides},}\ }\href@noop {} {\bibfield  {journal}
  {\bibinfo  {journal} {New J. Phys.}\ }\textbf {\bibinfo {volume} {21}},\
  \bibinfo {pages} {025004} (\bibinfo {year} {2019})}\BibitemShut {NoStop}%
\bibitem [{\citenamefont {Hotter}\ \emph {et~al.}(2020)\citenamefont {Hotter},
  \citenamefont {Plankensteiner},\ and\ \citenamefont {Ritsch}}]{HPR2020}%
  \BibitemOpen
  \bibfield  {author} {\bibinfo {author} {\bibfnamefont {C.}~\bibnamefont
  {Hotter}}, \bibinfo {author} {\bibfnamefont {D.}~\bibnamefont
  {Plankensteiner}}, \ and\ \bibinfo {author} {\bibfnamefont {H.}~\bibnamefont
  {Ritsch}},\ }\bibfield  {title} {\enquote {\bibinfo {title} {Continuous
  narrowband lasing with coherently driven v-level atoms},}\ }\href@noop {}
  {\bibfield  {journal} {\bibinfo  {journal} {New J. Phys.}\ }\textbf {\bibinfo
  {volume} {22}},\ \bibinfo {pages} {113021} (\bibinfo {year}
  {2020})}\BibitemShut {NoStop}%
\bibitem [{\citenamefont {S{\'a}nchez-Barquilla}\ \emph
  {et~al.}(2020)\citenamefont {S{\'a}nchez-Barquilla}, \citenamefont {Silva},\
  and\ \citenamefont {Feist}}]{SSF2020}%
  \BibitemOpen
  \bibfield  {author} {\bibinfo {author} {\bibfnamefont {M.}~\bibnamefont
  {S{\'a}nchez-Barquilla}}, \bibinfo {author} {\bibfnamefont {R.E.F.}\
  \bibnamefont {Silva}}, \ and\ \bibinfo {author} {\bibfnamefont
  {J.}~\bibnamefont {Feist}},\ }\bibfield  {title} {\enquote {\bibinfo {title}
  {Cumulant expansion for the treatment of light--matter interactions in
  arbitrary material structures},}\ }\href@noop {} {\bibfield  {journal}
  {\bibinfo  {journal} {J. Chem. Phys}\ }\textbf {\bibinfo {volume} {152}},\
  \bibinfo {pages} {034108} (\bibinfo {year} {2020})}\BibitemShut {NoStop}%
\bibitem [{\citenamefont {Jackson}(1999)}]{JDJ1999}%
  \BibitemOpen
  \bibfield  {author} {\bibinfo {author} {\bibfnamefont {J.~D.}\ \bibnamefont
  {Jackson}},\ }\href@noop {} {\emph {\bibinfo {title} {Classical
  Electrodynamics, 3rd Edition}}}\ (\bibinfo  {publisher} {John Wiley \&
  Sons},\ \bibinfo {year} {1999})\BibitemShut {NoStop}%
\bibitem [{dat()}]{data}%
  \BibitemOpen
  \bibfield  {title} {\enquote {\bibinfo {title} {Data for: {I}ntensity effects
  of light coupling to one- or two-atom arrays of infinite extent},}\
  }\href@noop {} {\bibinfo  {journal} {https://doi.org/10.4231/NWKJ-4B32}\
  }\BibitemShut {NoStop}%
\end{thebibliography}%

\end{document}